\documentclass[chicago]{emulateapj-rtx4}
\usepackage{amsmath}

\makeatletter
\newcommand\an{\ref@jnl{Astron. Nachtr.}}
\newcommand\rma{\ref@jnl{Rev. Mod. Astron.}}
\makeatother

\shorttitle{TESS in the Solar System}
\shortauthors{P\'al, Moln\'ar \& Kiss}

\begin{document}
\sloppy

\title{TESS in the Solar System}


\author{Andr\'as P\'al\altaffilmark{1,2,3}}
\author{L\'aszl\'o Moln\'ar\altaffilmark{1,4}}
\author{Csaba Kiss\altaffilmark{1}}
\altaffiltext{1}{Konkoly Observatory, Research Centre for Astronomy and Earth Sciences, Hungarian Academy of Sciences,
        Konkoly Thege Mikl\'os \'ut 15-17,
        H-1121 Budapest, Hungary}
\altaffiltext{2}{MIT Kavli Institute for Astrophysics and Space Research,
        70 Vassar St, Cambridge, MA 02109, USA}
\altaffiltext{3}{Department of Astronomy, Lor\'and E\"otv\"os University,
                 P\'azm\'any P. stny. 1/A,
                 Budapest H-1117, Hungary }
\altaffiltext{4}{MTA CSFK Lend\"ulet Near-Field Cosmology Research Group}
\email{apal@szofi.net}

\begin{abstract}
The Transiting Exoplanet Survey Satellite (TESS), launched successfully 
on 18th of April, 2018, will observe nearly the full sky and will 
provide time-series imaging data in $\sim$27-day-long campaigns. TESS is 
equipped with 4 cameras; each has a field-of-view of $24\times 24$ 
degrees. During the first two years of the primary mission, one of these 
cameras, Camera \#1, is going to observe fields centered at an ecliptic 
latitude of $18$ degrees. While the ecliptic plane itself is not covered 
during the primary mission, the characteristic scale height of the main 
asteroid belt and Kuiper belt implies that a significant amount of small 
solar system bodies will cross the field-of-view of this camera. Based 
on the comparison of the expected amount of information of TESS and 
Kepler/K2, we can compute the cumulative \'etendues of the two optical 
setups. This comparison results in roughly comparable optical 
\'etendues, however the net \'etendue is significantly larger in the 
case of TESS since all of the imaging data provided by the 30-minute 
cadence frames are downlinked rather than the pre-selected stamps of 
Kepler/K2. In addition, many principles of the data acquisition and 
optical setup are clearly different, including the level of confusing 
background sources, full-frame integration and cadence, the 
field-of-view centroid with respect to the apparent position of the Sun, 
as well as the differences in the duration of the campaigns. As one 
would expect, TESS will yield time-series photometry and hence 
rotational properties for only brighter objects, but in terms of spatial 
and phase space coverage, this sample will be more homogeneous and more 
complete. Here we review the main analogues and differences between the 
Kepler/K2 mission and the TESS mission, focusing on scientific 
implications and possible yields related to our Solar System.
\end{abstract}


\section{Introduction}
\label{sec:introduction}

The Transiting Exoplanet Survey Satellite (TESS) mission is an ongoing 
full-sky survey initiative, aiming to discover hundreds of rocky planets 
around main-sequence and dwarf stars \citep{ricker2015}. TESS orbits 
Earth in a special high inclination, high eccentricity orbit which has 
a period commensurable to the sidereal orbital period of the Moon, with 
a $2:1$ mean motion resonance. Onboard TESS, there are four wide-field 
cameras, each having $24\times 24$ degree of field-of-view (FOV) while 
the focal plane array is formed by $4$ frame-transfer CCDs with total 
dimensions of $2048\times 2048$ pixels. Hence, the pixel scale of TESS 
is $\sim21^{\prime\prime}$, however, due to the fast, $f/1.5$ focal 
ratio, this pixel scale varies slightly throughout the field-of-view. 
The cameras are aligned in a way that the gross FOV of TESS forms a 
nearly $96\times 24$ degrees wide strip in the sky.

\begin{figure}
\begin{center}
\plotone{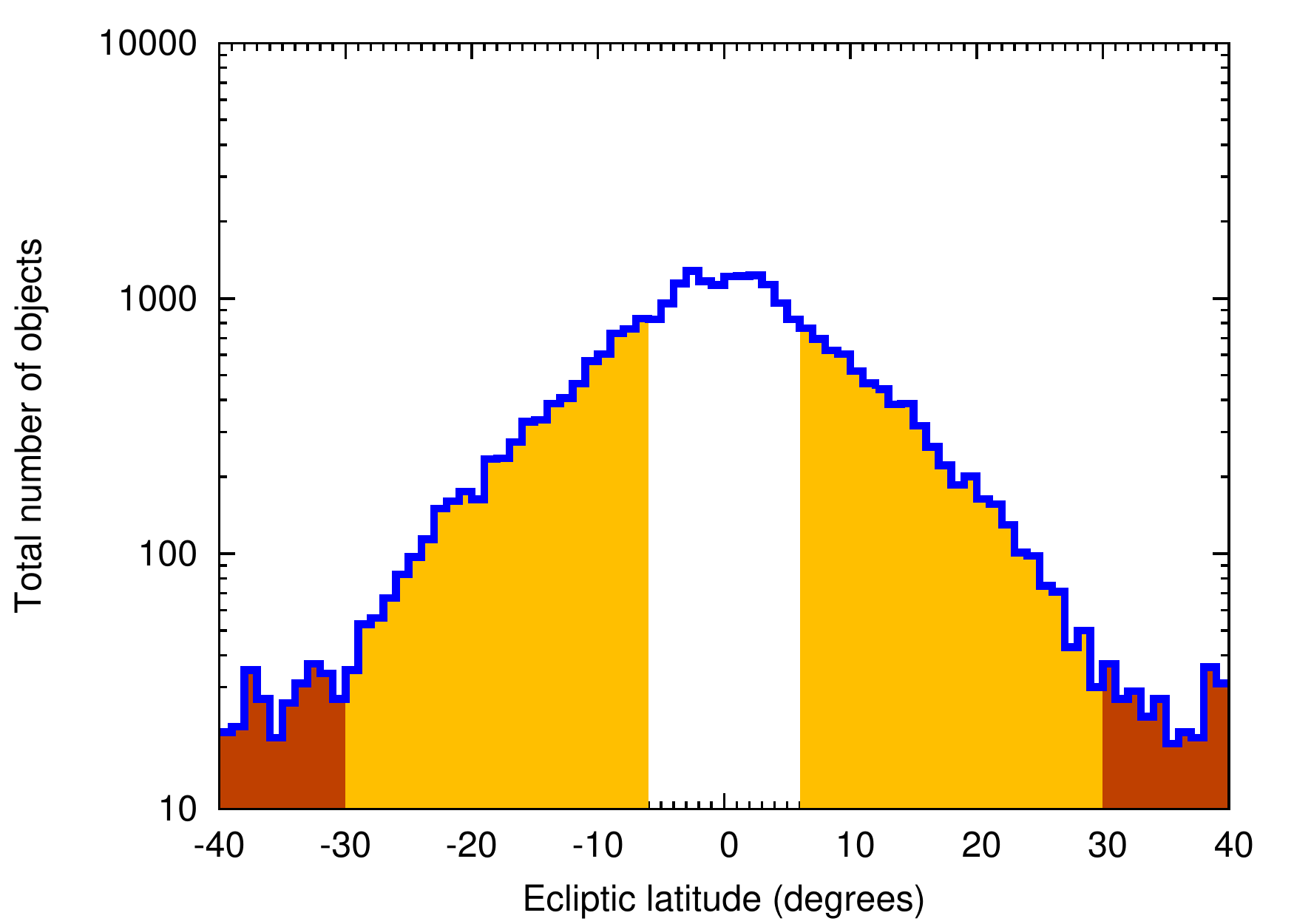}
\end{center}
\caption{The number density of the catalogued minor planets in a stripe 
having a width of 24 degrees as a function of the ecliptic latitude. 
The orange and brown filled parts correspond to TESS cameras \#4 and 
\#3, respectively. The empty part between $\pm 6$ degrees is not 
observed during the primary mission of TESS.}
\label{fig:segmentdrawing}
\end{figure}

Observations provided by TESS yield two main types of scientific data: 
one type of data are in the form of so-called \emph{stamps}, i.e. 
sub-frames with a 2-min cadence around several thousands of pre-selected 
F, G, K and M-type of stars. Another type of data are the full-frame 
images (FFI), which are provided with a cadence of 30 minutes. During
its two-year primary mission, TESS will observe $2\times 13$ fields, 
called \emph{sectors}, where its Camera \#4 points towards the ecliptic 
poles and Camera \#1 points close to the ecliptic plane, and the 
alignment of the cameras are parallel with ecliptic longitude great 
circles \citep[see Fig.~7 in][]{ricker2015}. There are $13$ partially 
overlapping sectors defined in the Northern Ecliptic Hemisphere and 
$13$, also partially overlapping sectors defined in the Southern 
Ecliptic Hemisphere \citep[see also Fig.~7 in][]{ricker2015}. Although 
this primary mission design avoids the ecliptic plane within a distance 
of $\sim 6^\circ$, the characteristic scale height of both the main 
asteroid belt and the Kuiper belt is significantly larger than this 
value -- namely, the meidan deviation and standard deviation of the 
inclinations are $\sim 7.1^\circ$ and $\sim 11.1^\circ$ 
for the main belt while $\sim 9.8^\circ$ and $\sim 24.0^\circ$ for
the Kuiper belt, respectively, based on the \texttt{MPCORB.dat.gz}
file (see Sec.~\ref{sec:tools}).
Therefore one can expect that a considerable number of moving 
objects will be detected by the TESS cameras. Indeed, even a simple 
query to an up-to-date minor planet database shows that a huge fraction 
of such bodies will be presented in the TESS FOV(s), mainly in Camera 
\#1 (and a few in Camera \#2), see also Fig.~\ref{fig:segmentdrawing}.

Our goal here is to provide an initial estimate on the TESS yield of 
photometric data related to minor Solar System bodies. The analysis 
presented here makes a detailed comparison with the same type of yield 
of the K2 mission \citep{howell2014}, which turned out to be a very 
effective instrument for targeted observations of these minor bodies 
\citep{szabo2015,pal2015}. The scientific highlights of the K2 mission 
include observations of a variety of Kuiper belt objects 
\citep{pal2015,pal2016a,benecchi2018}, natural satellites of gas giants 
\citep{kiss2016,farkas2017}, implications for the detection of the 
satellite of 2007 OR$_{10}$ \citep{kiss2017}, observations of main belt 
objects crossing K2 superstamps \citep{szabo2016,molnar2018} and 
observations of Jupiter Trojans \citep{szabo2017,ryan2017}. Extended and 
uninterrupted photometric coverage was found to be crucial to 
unambiguously identify long rotation periods that are commensurable with or 
longer than the day-night cycle of the Earth. Current asteroid rotation 
statistics have been found to be biased towards shorter periods by 
multiple studies \citep{masiero2009,marciniak2018,molnar2018}.

The structure of this paper goes as follows. First, in 
Sec.~\ref{sec:tessandk2} we briefly summarize the similarities and 
differences between the TESS and Kepler/K2 missions concerning solar 
system target observations. In Sec.~\ref{sec:tools} we describe the set 
of tools which are used in the TESS yield simulations in terms of 
observations of small bodies in our Solar System. In 
Sec.~\ref{sec:lcsimulations}, a set of examples are displayed by 
simulating 30-minute TESS FFI data using the up-to-date catalogues 
provided by the Minor Planet Center and the Gaia DR2 catalogue 
\citep{gaia2016,gaia2018} for background stars. These simulations 
focus on both the photometry of known objects, in order to retrieve 
rotation characteristics, as well as on the estimates of flux excess 
during the photometry of stars. In Sec.~\ref{sec:encounters}, statistics 
are presented for the expected flux excess of target stars caused by 
minor planet encounters. We summarize our results and conclude in 
Sec.~\ref{sec:summary}.

\section{Similarities and differences between TESS and Kepler/K2}
\label{sec:tessandk2}

After the failure of the second reaction wheel of the {\it Kepler} Space 
Telescope \citep{borucki2010}, the K2 mission was initiated by targeting 
{\it Kepler} to the ecliptic plane in order to minimize the required 
thruster firing corrections \citep{howell2014}. These comparatively 
frequent thruster firing corrections are necessary due to the torques 
induced by solar radiation pressure and introduces a larger amount of 
systematic noise in the raw light curves. Similar to the {\it Kepler} 
space telescope, TESS is also designed for extremely precise photometry 
of stars with fixed positions in the sky. In the following, we list the 
main similarities and differences of the Kepler/K2 and TESS data 
acquisition schemes, which will also affect the selection of the most 
suitable targets within our Solar System.

\begin{figure}
\begin{center}
\resizebox{70mm}{!}{\includegraphics{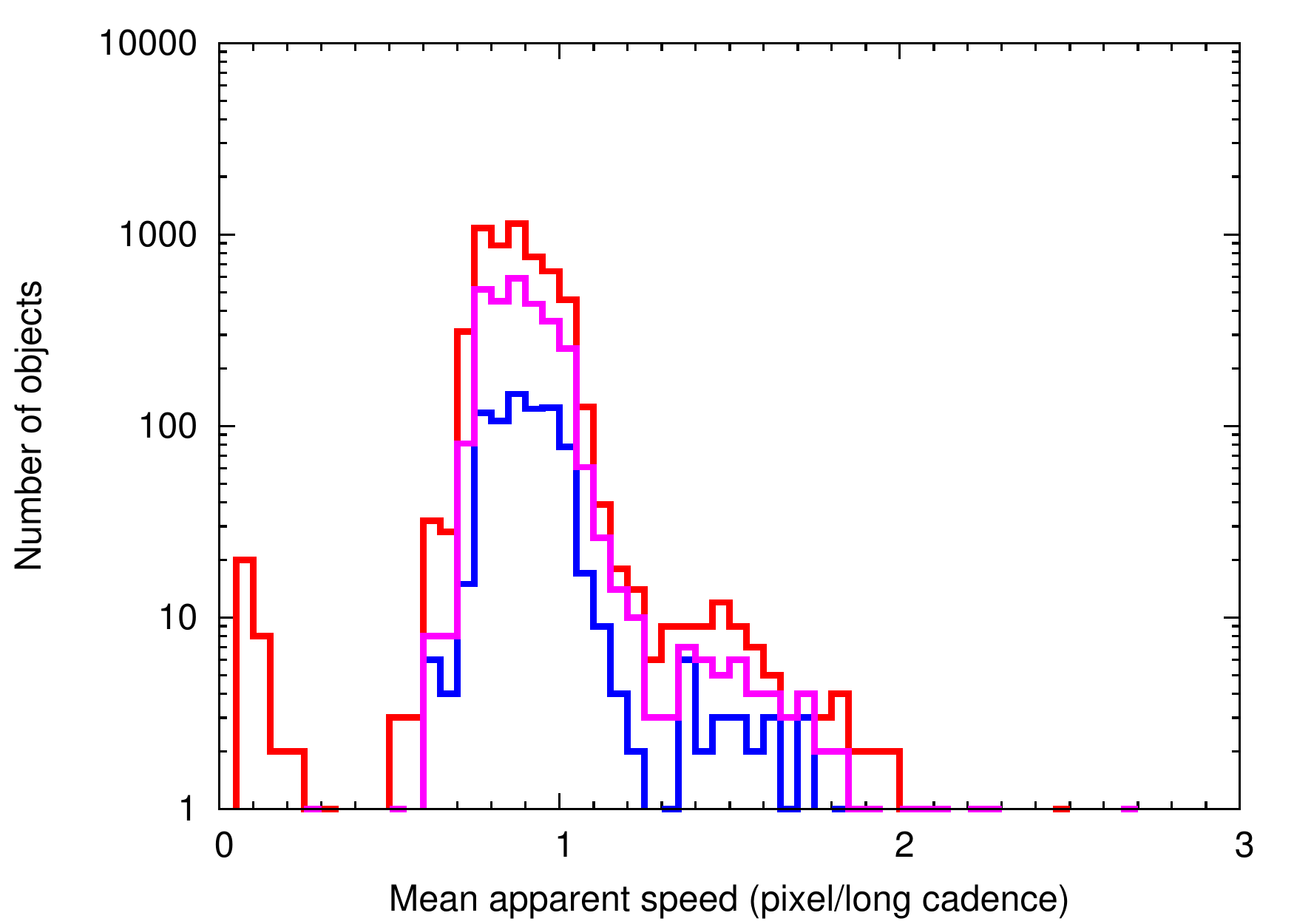}}
\end{center}
\caption{The number density of the minor planets (Camera \#1, closest to 
the ecliptic plane) as a function of the apparent speed. The red curve 
shows this number density for all of the catalogued objects, while the 
purple and blue curves correspond to the objects brighter than $V=20$ 
and $V=18.5$ magnitudes, respectively. The minor planets have been 
queried for a single TESS campaign.}
\label{fig:shist}
\end{figure}

First, while the typical apparent speed of a main-belt asteroid is $3-5$ 
pixels per $30$-min long cadence for K2 \citep{szabo2016,molnar2018}, it is 
significantly smaller in the case of TESS, i.e., it is smaller or close 
to 1\,pixel/cadence ($\lesssim 21^{\prime\prime}/{\rm 30\,mins}$, see 
also Fig.~\ref{fig:shist}). Therefore, we can make reliable estimates 
for TESS just considering non-moving targets at first glance -- see, 
e.g., Fig.~14 in \cite{sullivan2015} or Fig.~8 in \cite{ricker2015}.

\begin{figure*}
\begin{center}
\resizebox{35mm}{!}{\includegraphics{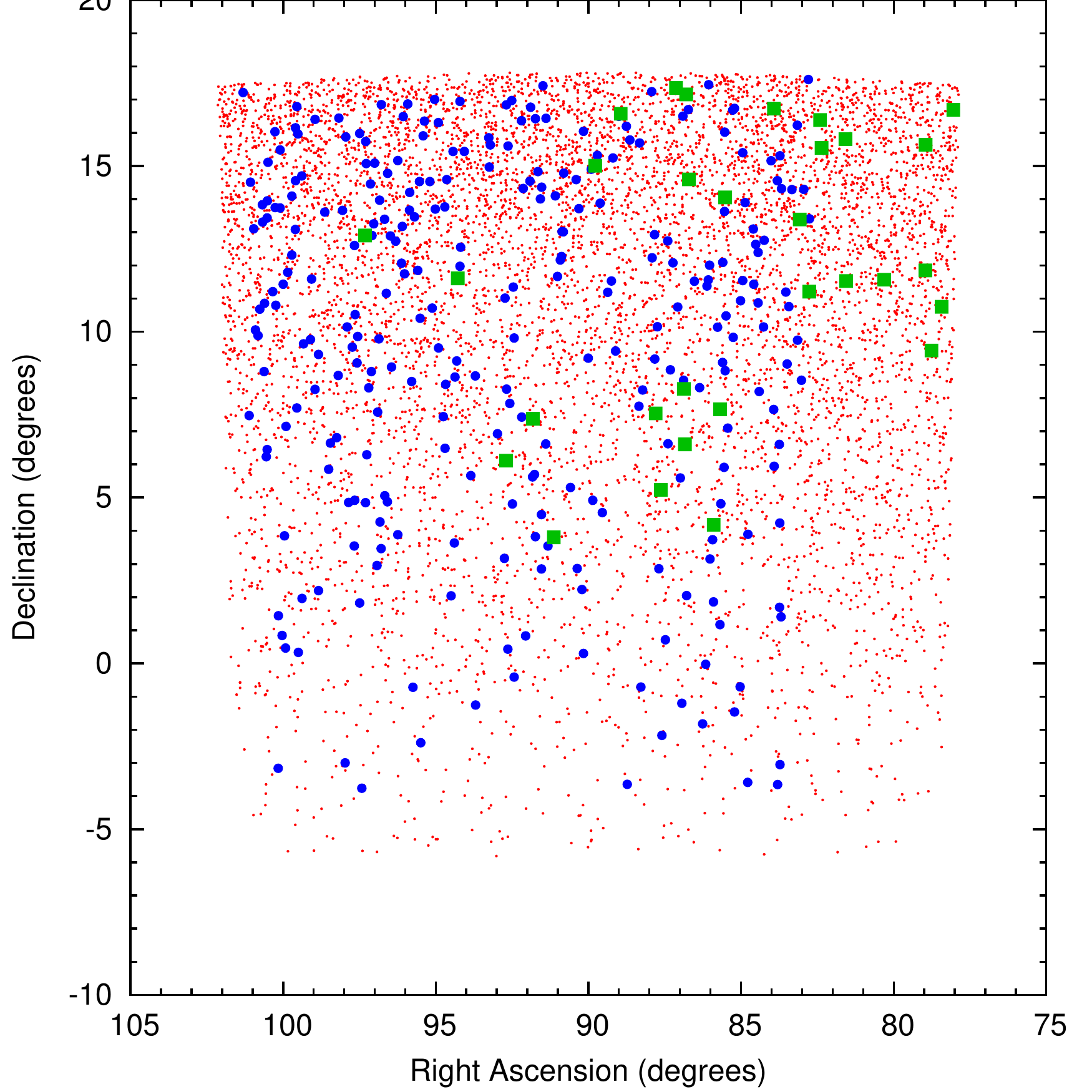}}%
\resizebox{35mm}{!}{\includegraphics{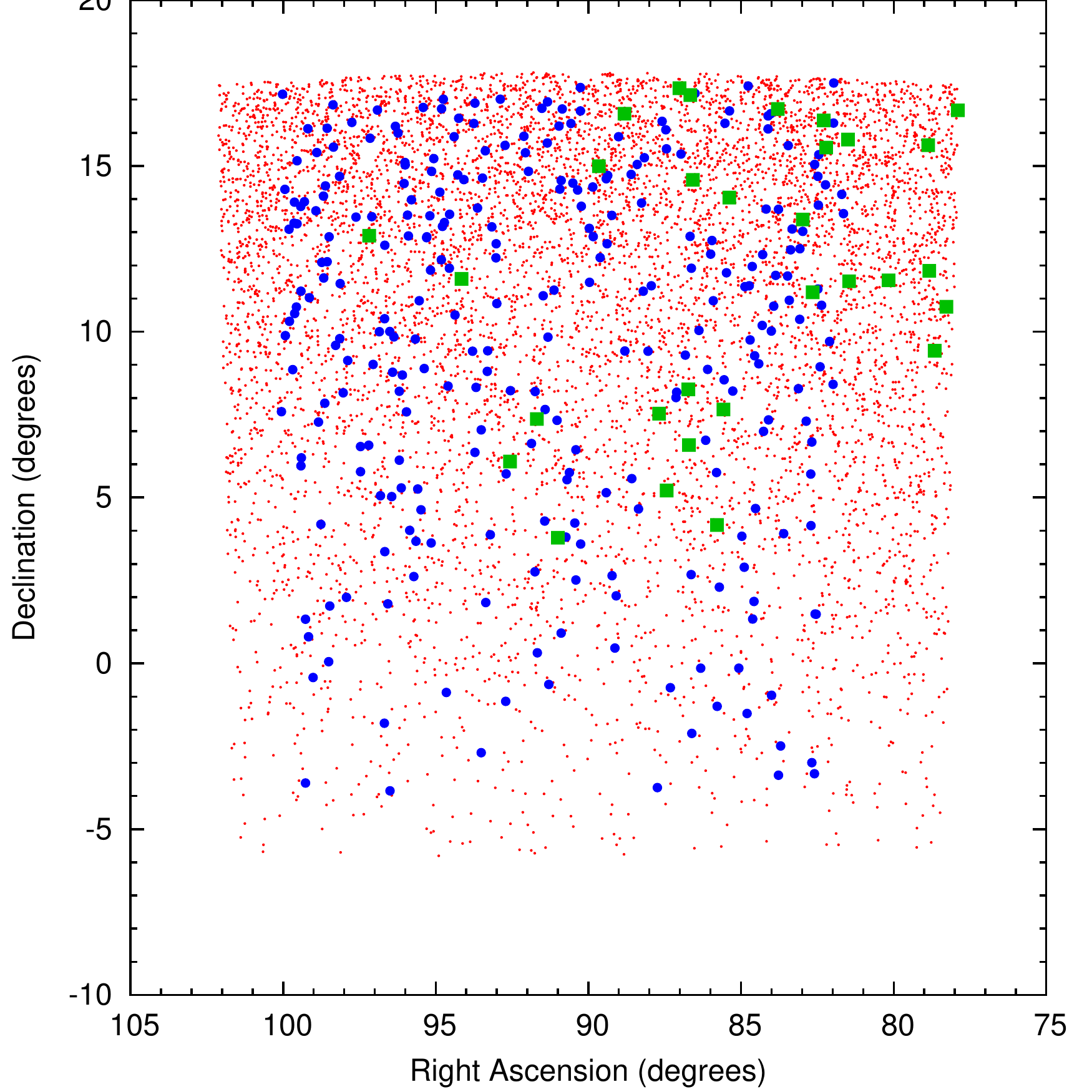}}%
\resizebox{35mm}{!}{\includegraphics{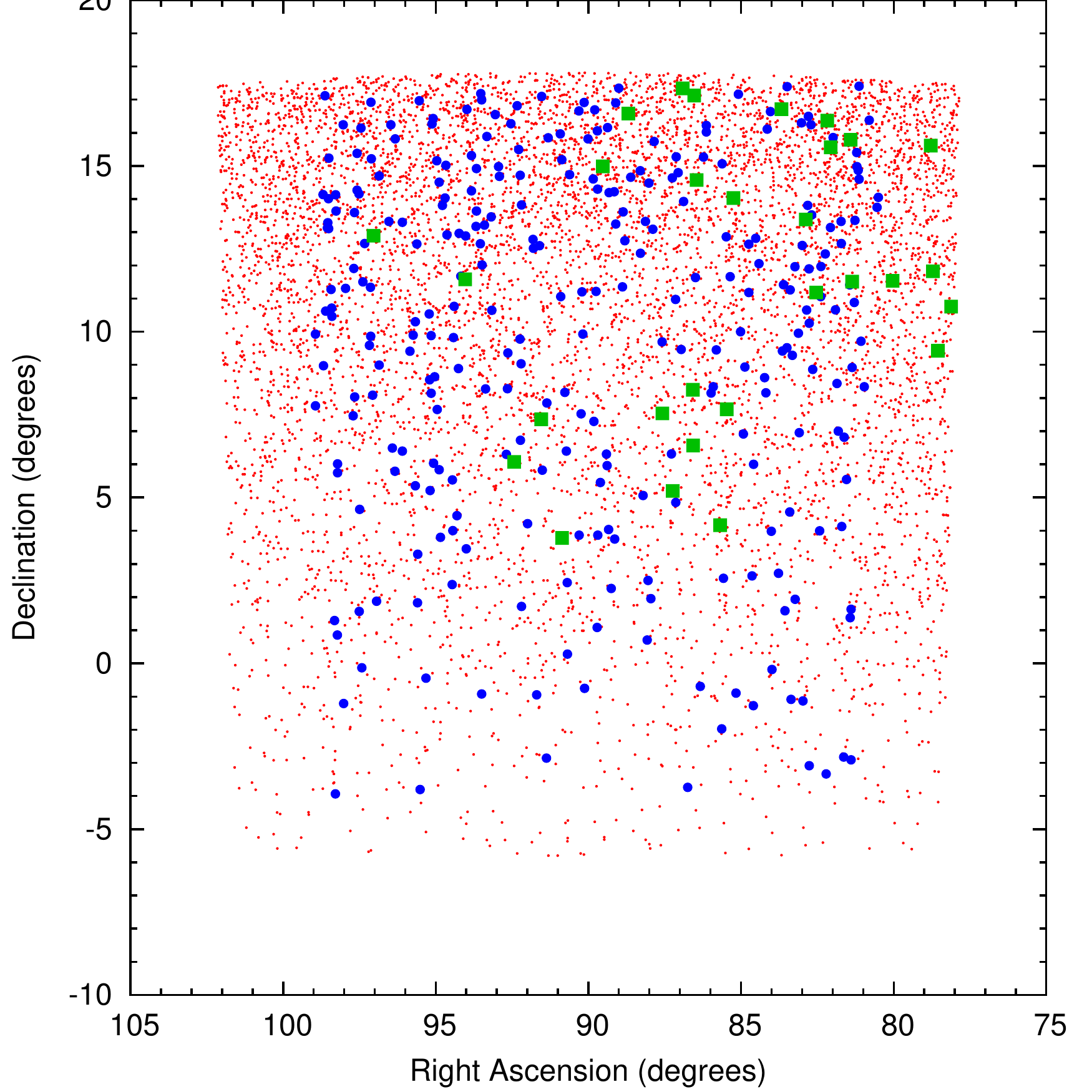}}%
\resizebox{35mm}{!}{\includegraphics{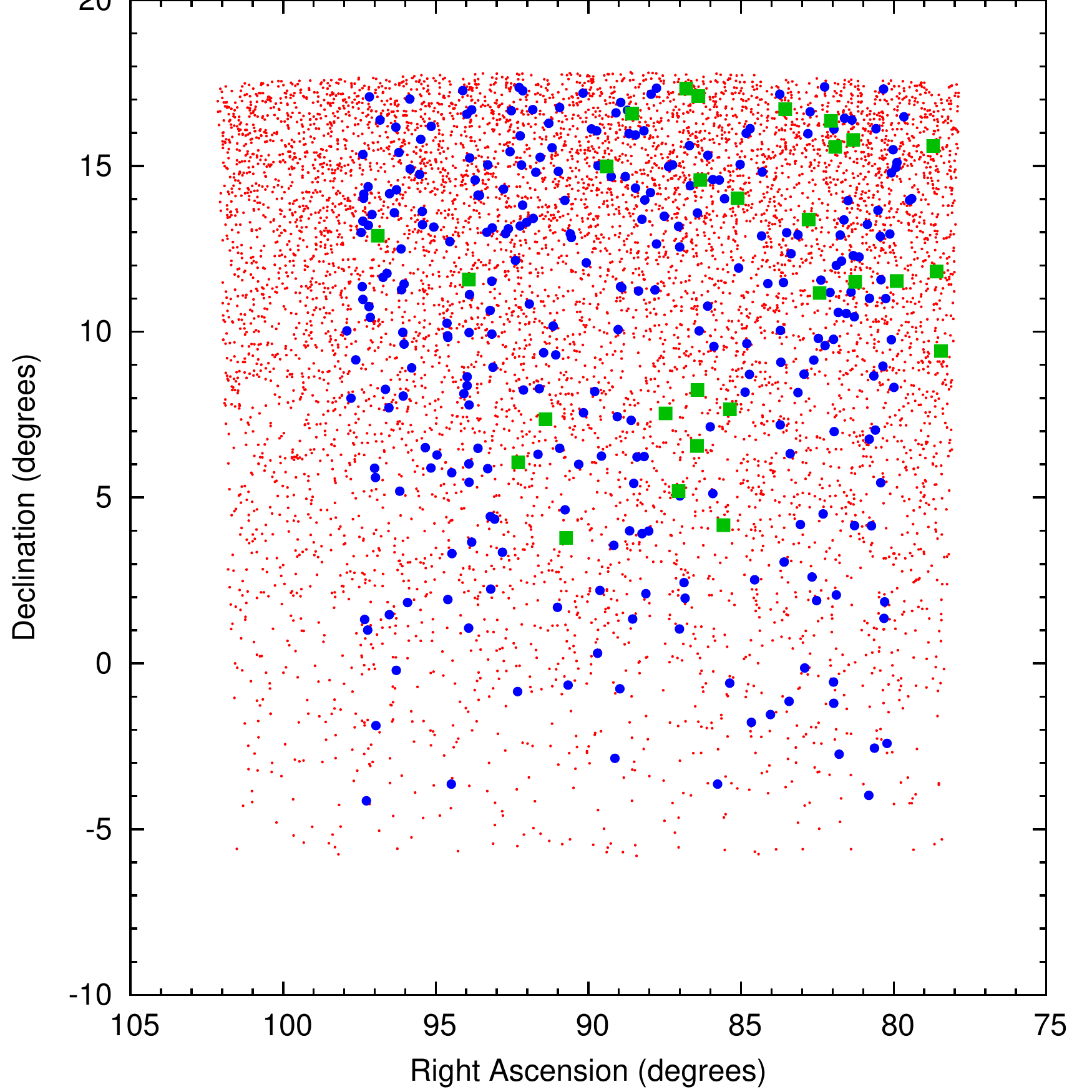}}%
\resizebox{35mm}{!}{\includegraphics{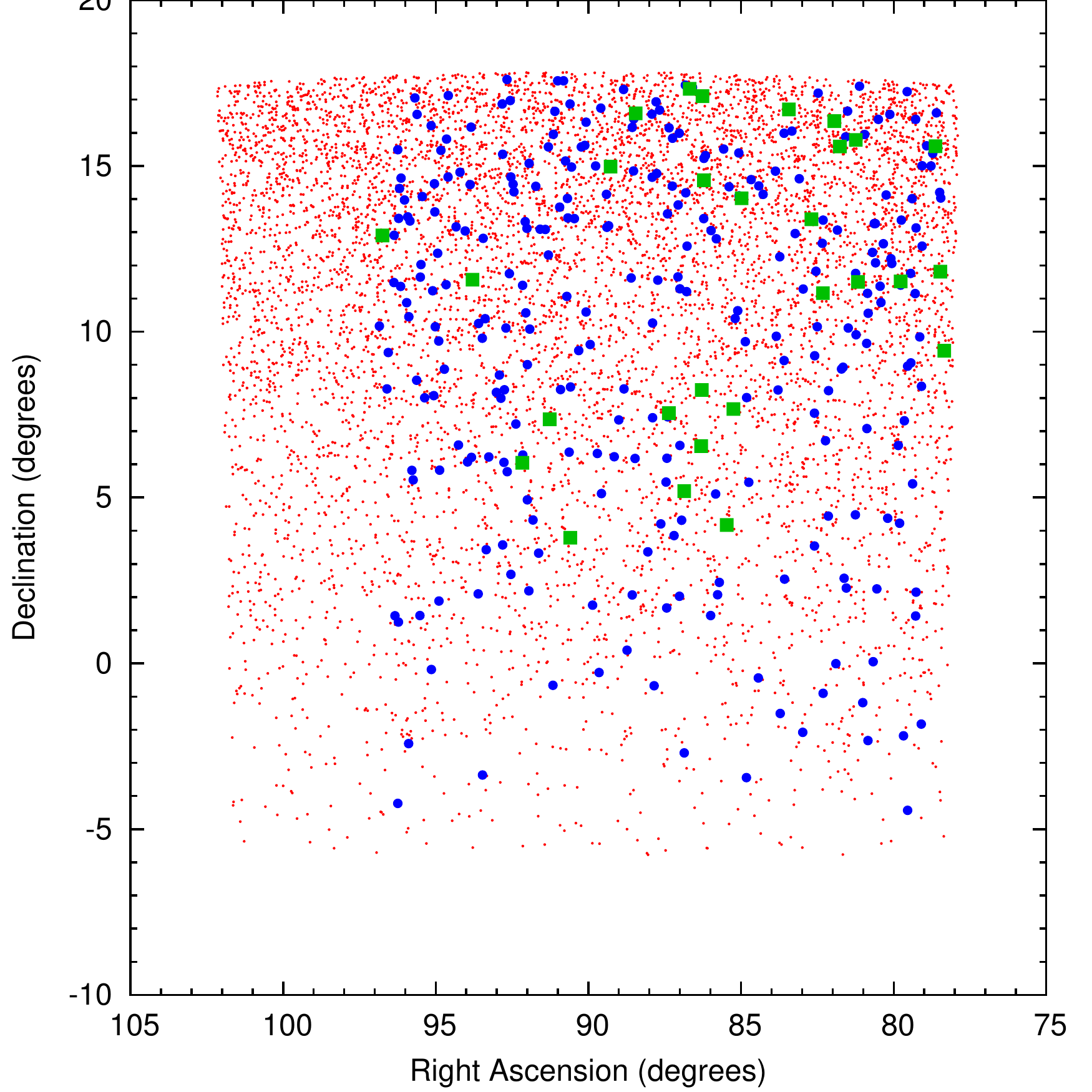}}
\end{center}
\caption{The series of images above shows all catalogued minor planets 
(as of 2017 June 12) as it is seen by Camera \#1, which is the 
closest to the ecliptic plane. The cadence between these images are 5 
days while the ephemerides of the central image correspond to JD~2458474 
(2018 December 21, 12:00 UTC). The small red dots correspond to the 
catalogued minor planets. For visualization purposes, the blue dots 
mark every 20th object that can be seen during the whole campaign. In 
addition, green squares show the positions of trans-Neptunian objects 
(TNOs). The swarm of the blue dots clearly show that during a one-month 
TESS campaign effectively all minor planets have a correlated and 
retrograde motion. Note that only those small bodies that appear in the 
first panel are displayed throughout the series of figures.}
\label{fig:camsupport}
\end{figure*}

Second, in the case of TESS, 30--50\% of the objects that cross the 
field-of-view can be seen throughout the whole one-month campaign. This 
can easily be computed from the previously mentioned estimates of 
$\lesssim 1\,{\rm pixel}/{\rm cadence}$ for the apparent speed, the 
total number of FFI cadences acquired during the observations of a 
single sector (which is around $\sim 1300$ for a 27-day-long campaign 
per sector) and the width of the sectors (i.e., the size of the sectors 
parallel with the ecliptic latitude circles) which is in the range of 
$4000$ pixels in total. A series of 5 images displaying this behavior is 
presented in Fig.~\ref{fig:camsupport}.

In addition, in the case of TESS, minor planets are observed during 
opposition while K2 observes these closer to their stationary points. 
Therefore the apparent movement of the objects are correlated, 
all retrograde, and the velocity dispersion is much smaller than for K2
(see also Figs.~\ref{fig:shist} and \ref{fig:camsupport}).

In order to further quantify our expectations about the usefulness of 
TESS for solar system science, we ran simulations by involving some of 
our tools. The main concepts of these simulations and the tools employed 
therein are described in the following section (Sec.~\ref{sec:tools}) 
while the resulting light curves are displayed and detailed in 
Sec.~\ref{sec:lcsimulations}.

\section{Tools for data reduction, analysis and astrometric predictions}
\label{sec:tools}

In order to get an estimate on the precision and accuracy of 
Solar System observations, we carried out a 
series of simulations in a Monte-Carlo fashion. Namely, artificial 
images were generated which have similar characteristics to TESS, 
considering both the optical and imaging properties as well as the 
detector parameters such as readout and other types of noise. This 
series of simulations was build on the top of four databases and 
software packages as follows:

\begin{itemize}
\item \texttt{MPCORB.dat.gz} -- the \texttt{MPCORB.dat.gz} file provided 
by the Minor Planet Center. This is a simple textual file with several 
hundreds of thousands of lines, containing orbital elements and 
brightness parameters (absolute magnitudes and slope parameters) for all 
of the catalogued minor planets in the Solar System. This file is 
regulary updated and can be retrieved from the Minor Planet 
Center\footnote{http://www.minorplanetcenter.net/iau/MPCORB/MPCORB.DAT.gz}. 
Although online tools (such as MPC 
itself\footnote{http://www.minorplanetcenter.org/iau/mpc.html} and 
JPL/Horizons\footnote{https://ssd.jpl.nasa.gov/horizons.cgi}) exist 
that can retrieve a series of ephemerides for single or multiple 
objects, our simulations require the nearly simultaneous handling of all 
of the available orbital data of minor planets. Therefore, it is easier 
and safer to handle all of the related computations in an offline 
fashion since both the number of epochs and the size of the 
field-of-view(s) are much larger than the intended use-cases of these 
available online tools. 
\item \textit{Gaia} DR2 -- the recently 
published \textit{Gaia} DR2 catalogue. This catalogue 
\citep{gaia2016,gaia2018} contains records for $\sim 1.6$ billion point 
sources. Since the wide-band TESS throughput \citep[see Fig.~1 
in][]{ricker2015} quite well agrees with the \textit{Gaia} 
$G_\mathrm{RP}$ band \citep[see Fig.~3 in][]{jordi2010}, this database 
is rather suitable to simulate the stellar background of the simulated 
images by employing the $G_\mathrm{RP}$ magnitudes (i.e. the 61st, the 
\texttt{phot\_rp\_mean\_mag} record in the \texttt{*.csv.gz} files in 
the \texttt{./gdr2/gaia\_source/csv} directory, see also the original 
database 
descriptions\footnote{https://www.cosmos.esa.int/web/gaia/dr2}). 
\item EPHEMD. While several tools exist which can be used as ephemeris 
services, none of them are optimized for retrieving a list of Solar 
System objects that traverse the large field-of-view of TESS. This 
feature of a service is generally referred as \emph{cone search}. The 
web-based service of the Minor Planet Center is suitable only for 
ground-based observatories while the Virtual Observatory service SkyBoT 
\citep{berthier2016} is capable of identifying minor planets only for a 
single time instance. While developed mainly for the observations 
related to the K2 mission \citep{molnar2018}, here we list some of the 
features of the tool named EPHEMD, optimized for searches spanning 
longer time intervals -- hence, this tool is also optimal for 
TESS. \begin{itemize} \setlength{\itemsep}{0pt} \item EPHEMD is 
implemented in a server-client architecture. The program itself is 
running in the background while it caches all of the necessary 
pre-computed orbits. Simple commands referring to queries and ephemeris 
services can then be executed while connecting via TCP/IP. \item The 
input database of EPHEMD is derived from the file 
\texttt{MPCORB.dat.gz}. 
updated by the Minor Planet Center. \item EPHEMD supports easy 
integration of various observers (including the Kepler spacecraft and 
TESS) by involving either SPICE kernels or ephemerides for the 
observers. \item The precision of the computations performed by EPHEMD 
is in the range of $\lesssim 0.02^{\prime\prime}$ with respect to the 
JPL/Horizons service (namely, differences can only be seen in the last 
significant digit when one queries positions for 5 decimal places in 
degrees). \end{itemize} A more detailed description of this EPHEMD 
package will be available soon along with the uploading of its source 
code to the public domain as a free and open source program. 
\item FITSH. The FITSH package \citep{pal2012} is used as the core of a 
state-of-the-art pipeline developed for the photometry of Solar System 
objects in the fields of Kepler/K2 missions \citep{pal2015,szabo2016}. 
Features that are relevant for analyzing this type of data include the 
elongated apertures supported by FITSH/\texttt{fiphot}. In addition, the 
tasks of FITSH perform astrometric analysis and automated cross-matching 
of sources, including the computations corresponding to the large 
optical distortions of the TESS cameras. While this feature is not 
essential during the simulations, analysis of real data needs such an 
algorithm, similarly to the way it is employed in the case of K2 
superstamps \citep[see, for instance][]{farkas2017,molnar2018}. 
\end{itemize}

\begin{figure*}
\begin{center}
\noindent%
\resizebox{70mm}{!}{\includegraphics{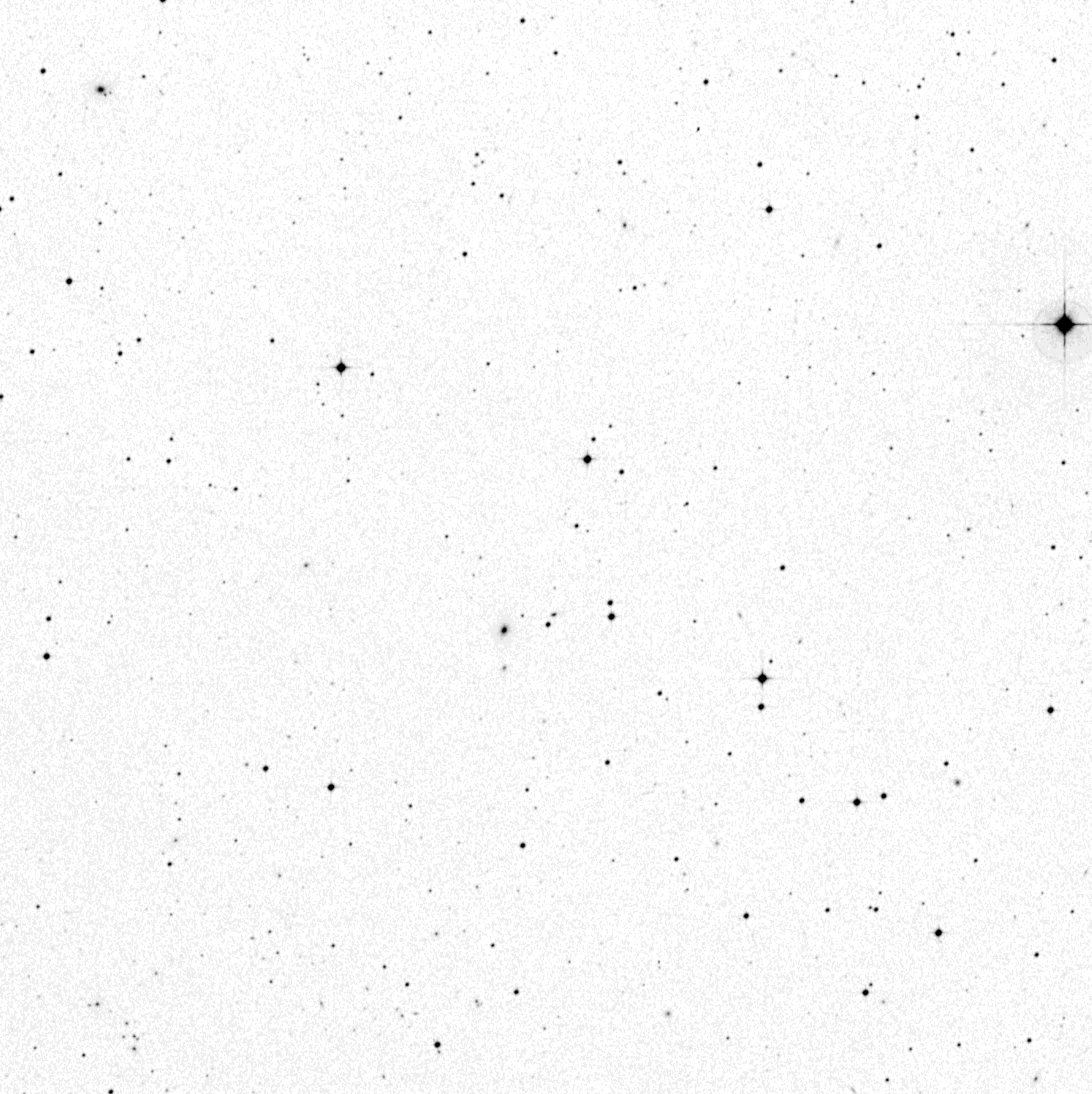}}\hspace*{10mm}%
\resizebox{70mm}{!}{\includegraphics{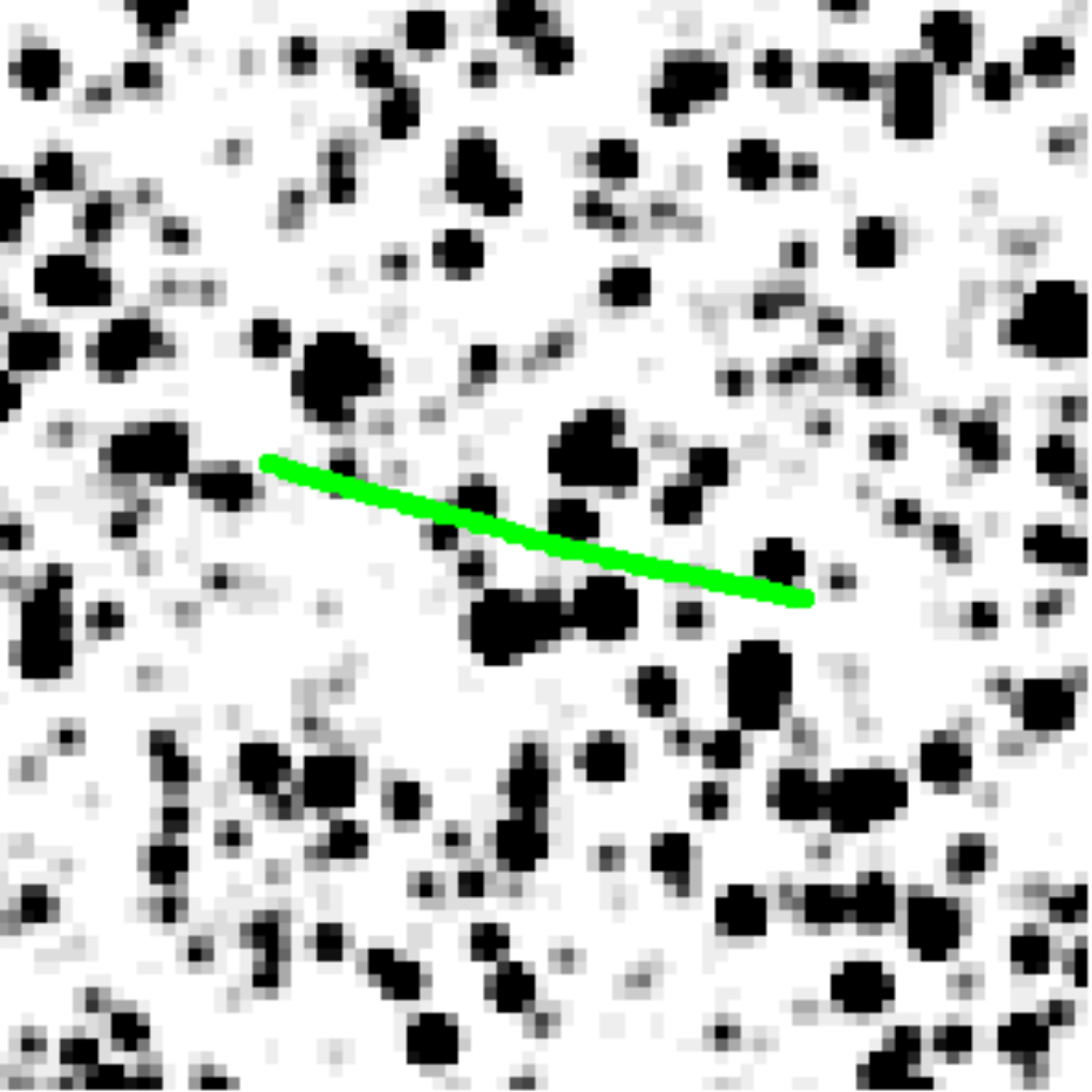}}\vspace*{4mm}

\noindent%
\resizebox{80mm}{!}{\includegraphics{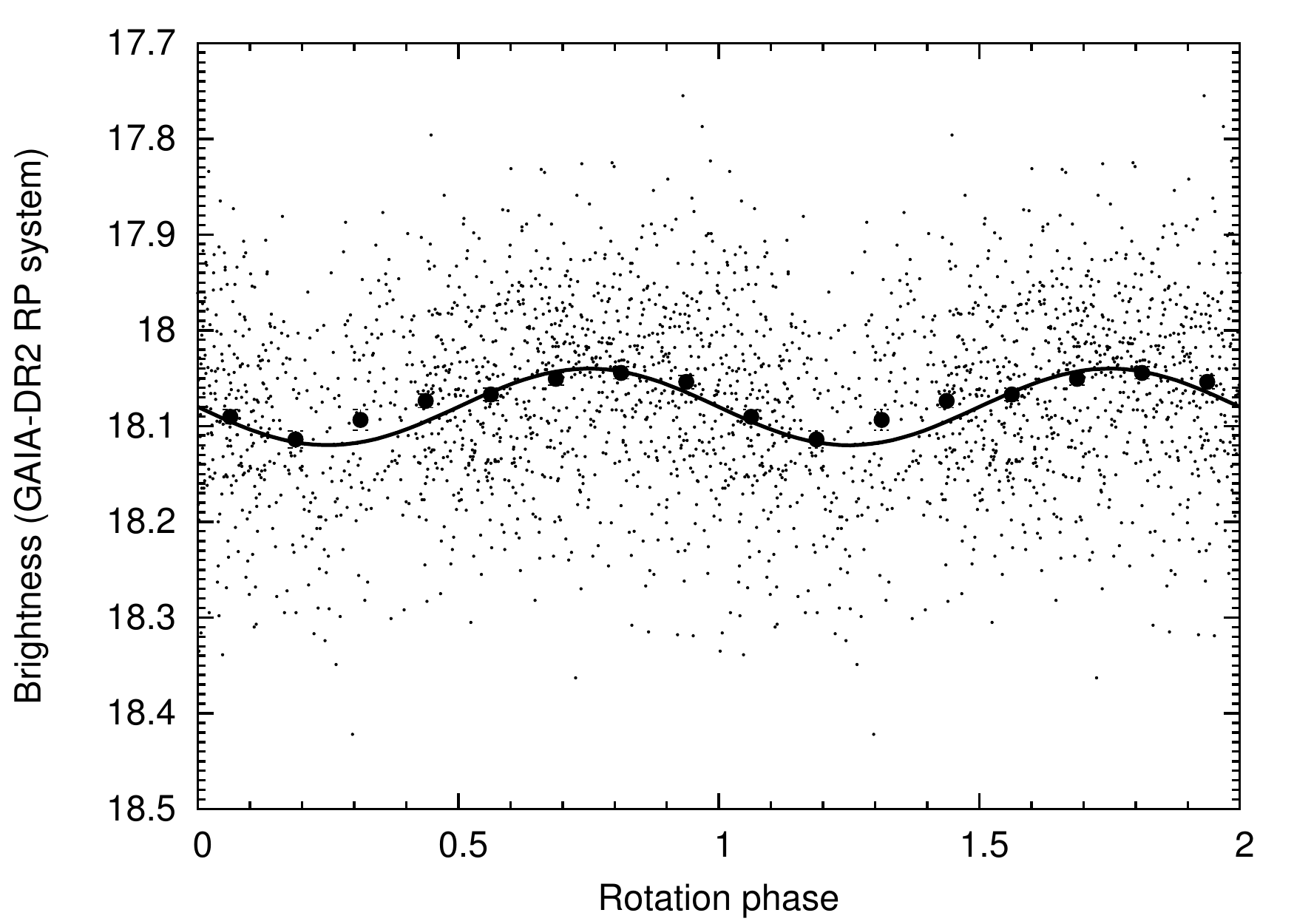}}%
\resizebox{80mm}{!}{\includegraphics{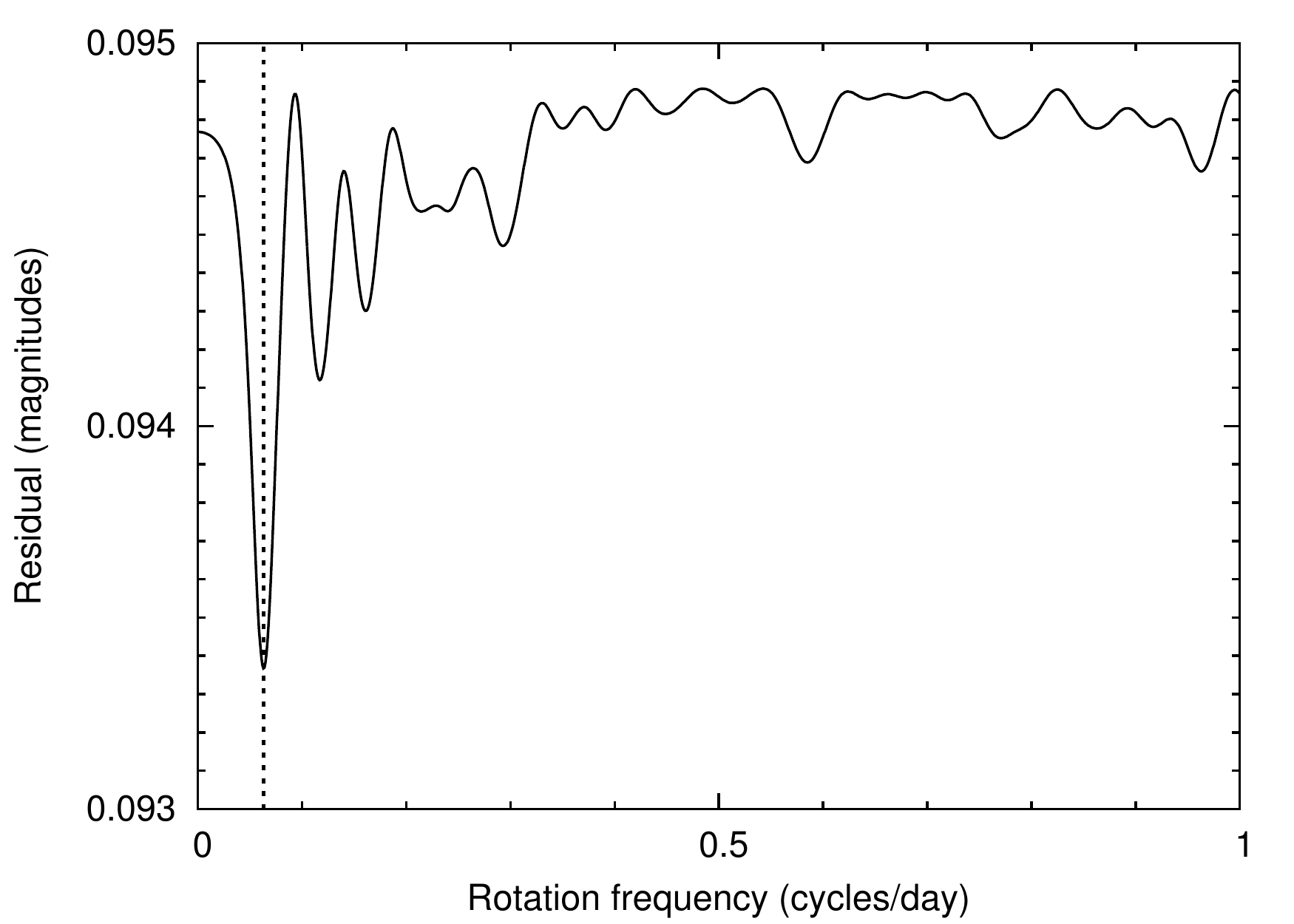}}
\end{center}
\caption{Simulations of the TESS observations for the dwarf planet Eris.
\emph{Upper left:} The $30^{\prime}\times30^{\prime}$ field-of-view of
the sky, centered at $\alpha=25.842$ and $\delta=-2.262$. This position
corresponds to the apparent location of Eris on October 20, 2018. During the
first year of science operations, TESS will observe the Southern Hemisphere
where Eris is presently located. \emph{Upper right:} The same field-of-view
in the simulated TESS images. During this simulation, the alignment of TESS
cameras is set w.r.t.\ the equatorial reference system. The green line
shows the path of Eris as it traverses this field for 27 days. 
\emph{Lower left:} Reconstructed folded light curve of Eris, expecting a
single-peaked solution with a rotation period of $15.77\,{\rm d}$. The 
peak-to-peak amplitude of this injected signal is $0.05$ mags. 
\emph{Lower right:} Residual spectrum of Eris' light curve as a function
of the rotation frequency. Due to the single-peaked nature, this spectrum is
dominated by the peak corresponding to the injected rotation frequency.}
\label{fig:eris}
\end{figure*}

The actual simulations are carried out as follows. By retrieving the 
list of \textit{Gaia} DR2 sources as well as pre-computed ephemerides of 
minor bodies, an artificial image is created by the 
FITSH/\texttt{firandom} tool after applying the appropriate astrometric 
projections (which convert J2000 ICRS RA and DEC coordinates into the 
$x,y$ plane of the CCD detector). Shot noise and background noise is 
added in accordance with the expected signal-to-noise levels \citep[see 
Fig.~8 in ][]{ricker2015}.

\begin{figure*}
\begin{center}
\hfill
{\bf High galactic lat.:} $|\beta|\approx 45^\circ$~~~~~~~
\hfill
{\bf Medium galactic lat.:} $|\beta|\approx 20^\circ$~~~~
\hfill
{\bf Low galactic lat.:} $|\beta|\approx 10^\circ$
\hfill\,

\noindent
\resizebox{57mm}{!}{\includegraphics{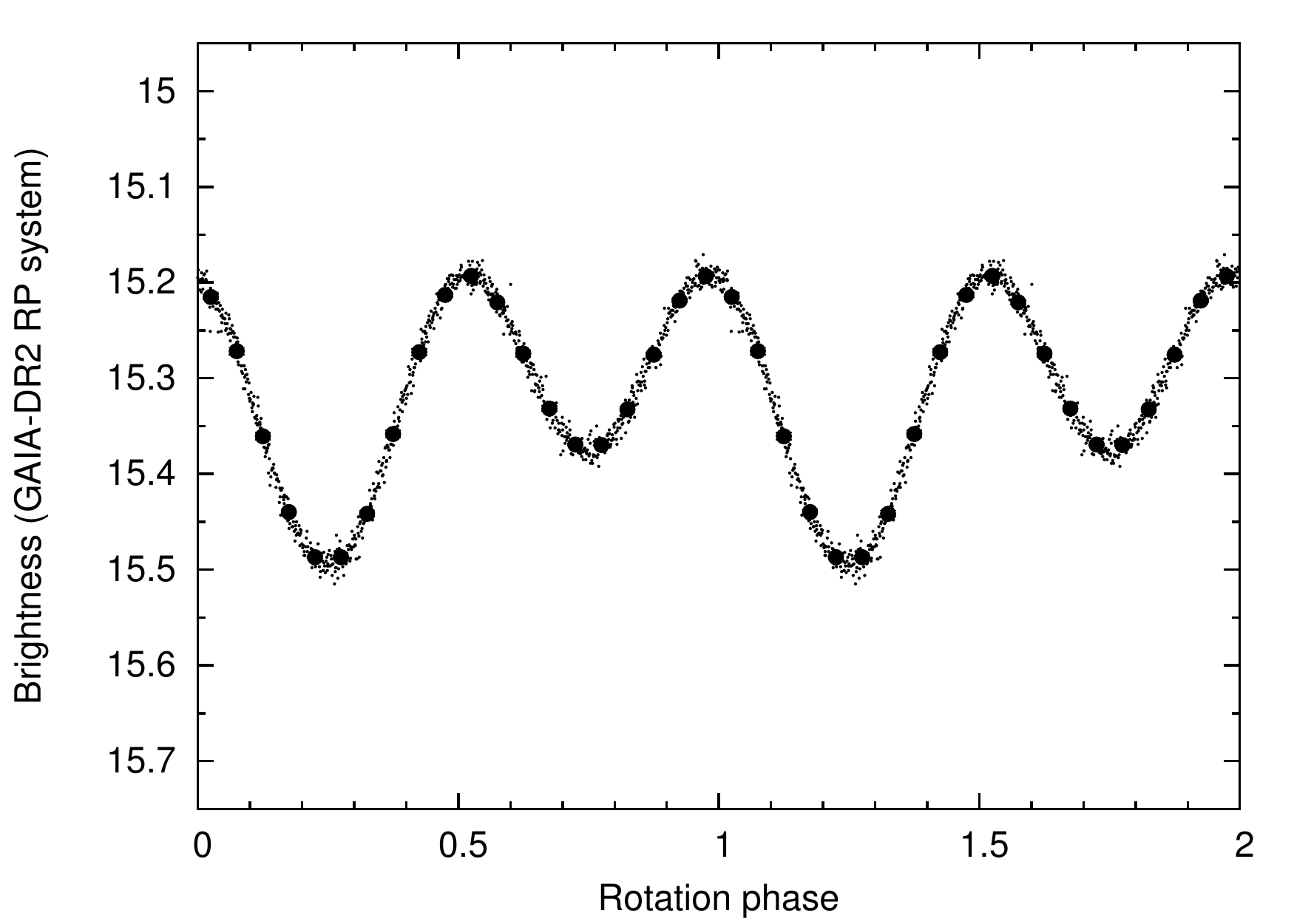}}%
\resizebox{57mm}{!}{\includegraphics{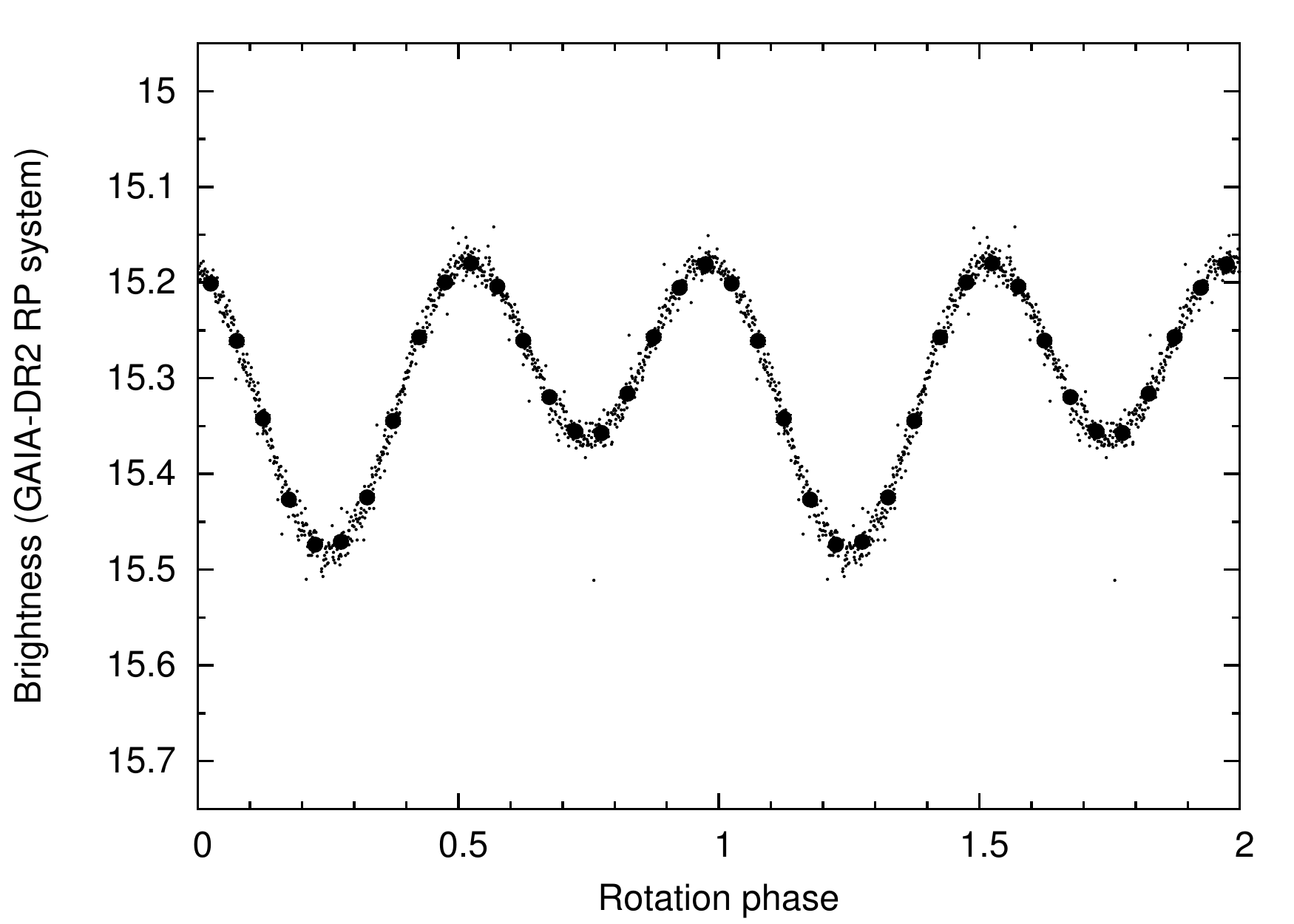}}%
\resizebox{57mm}{!}{\includegraphics{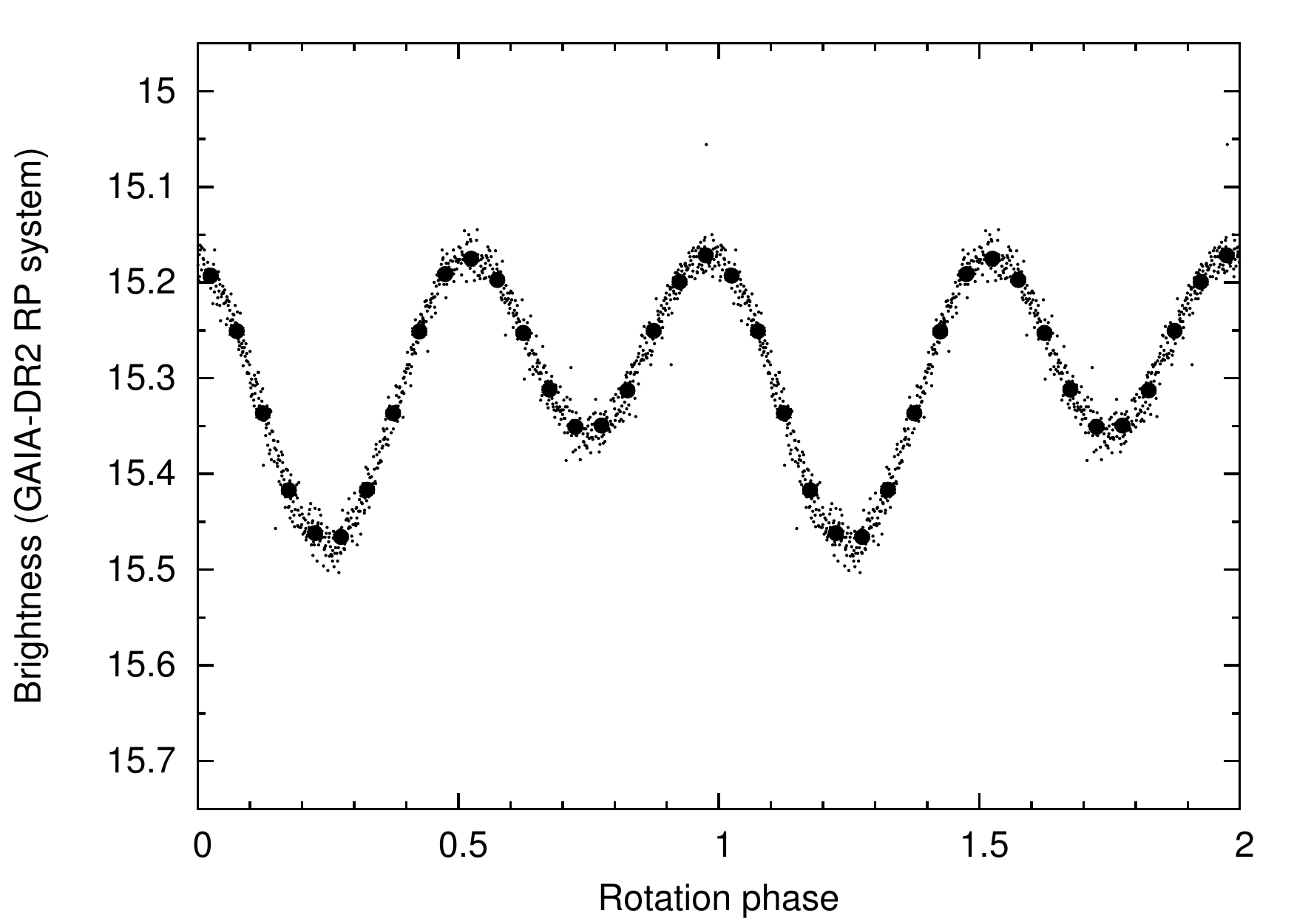}}%

\noindent
\resizebox{57mm}{!}{\includegraphics{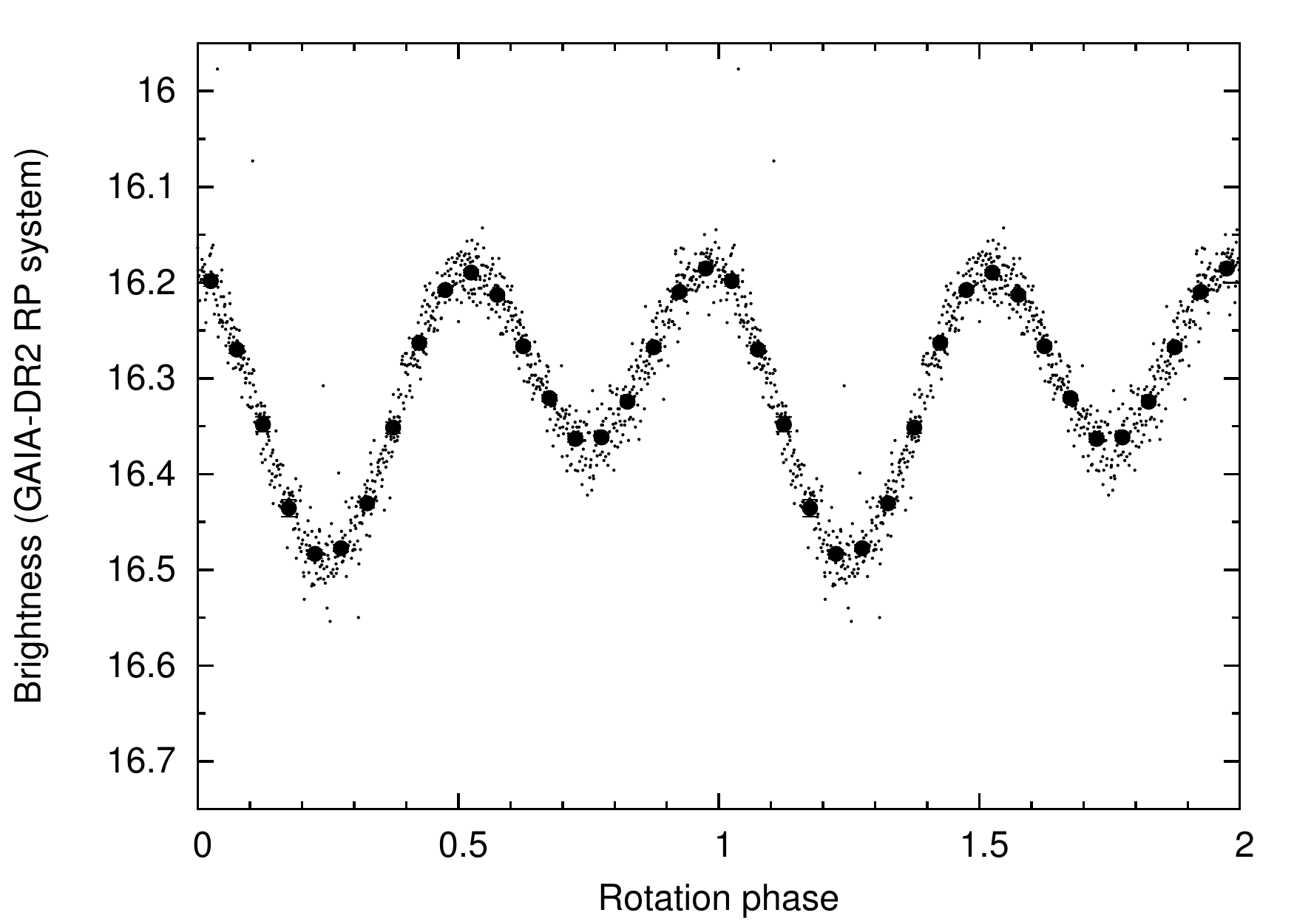}}%
\resizebox{57mm}{!}{\includegraphics{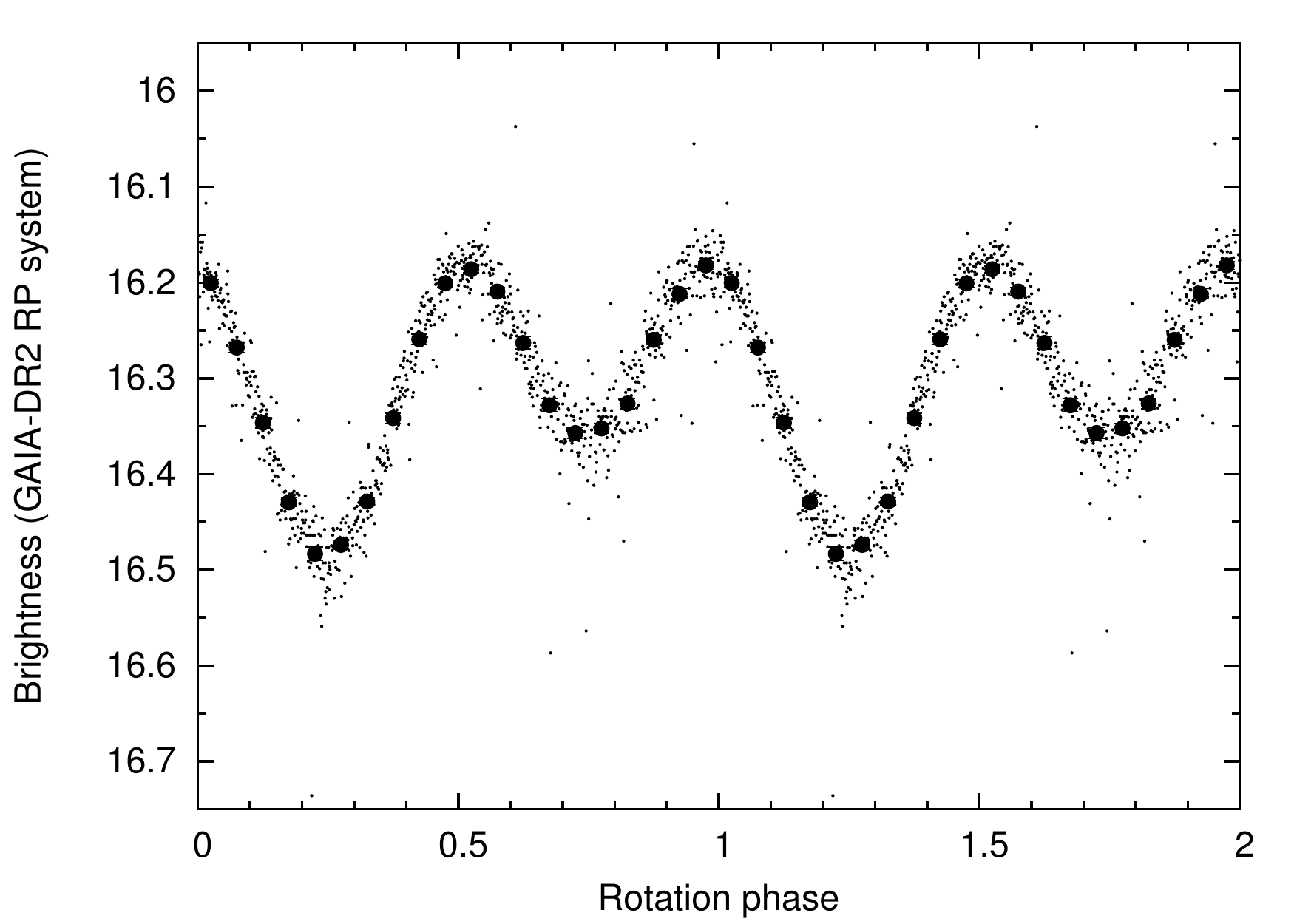}}%
\resizebox{57mm}{!}{\includegraphics{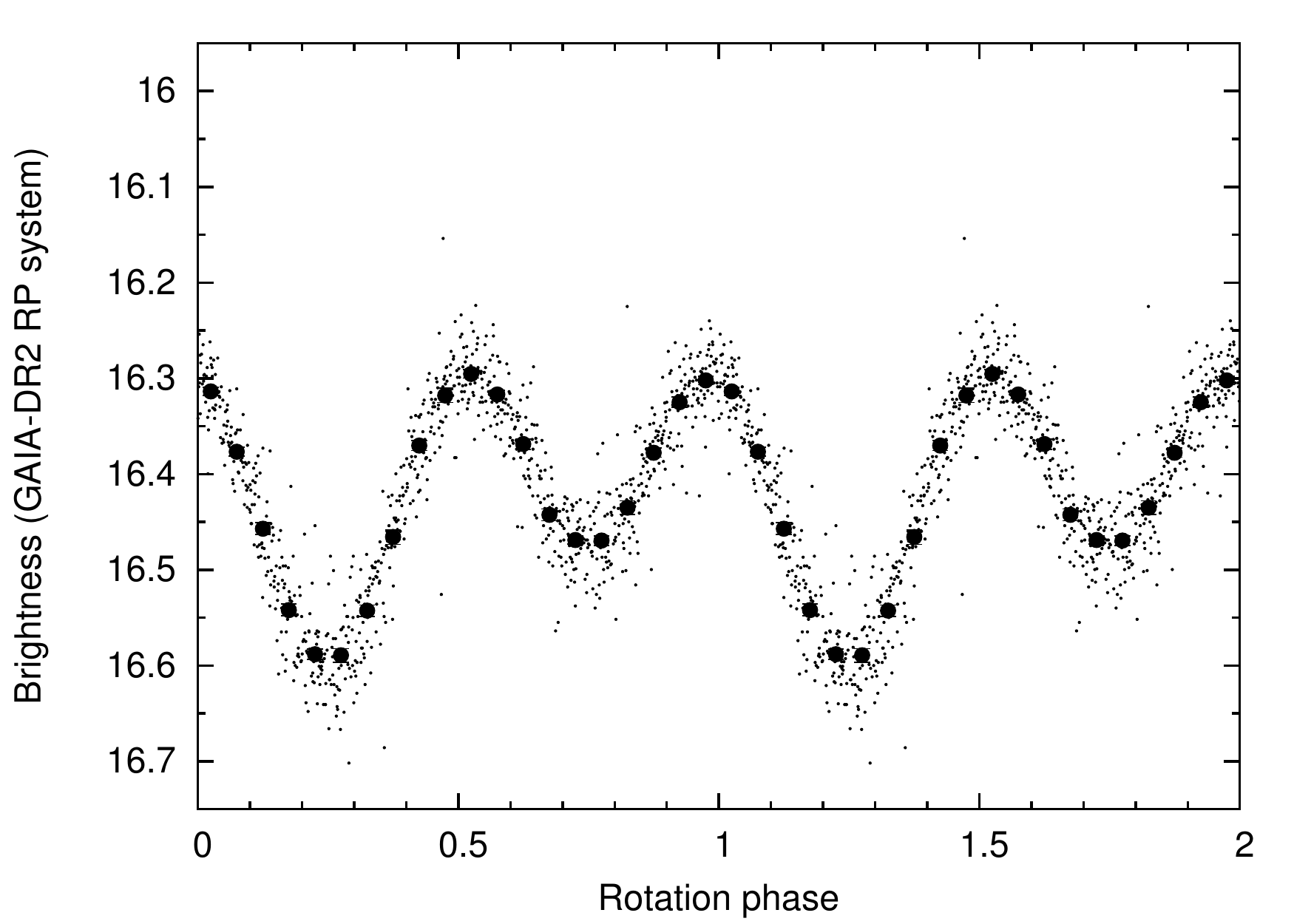}}%

\noindent
\resizebox{57mm}{!}{\includegraphics{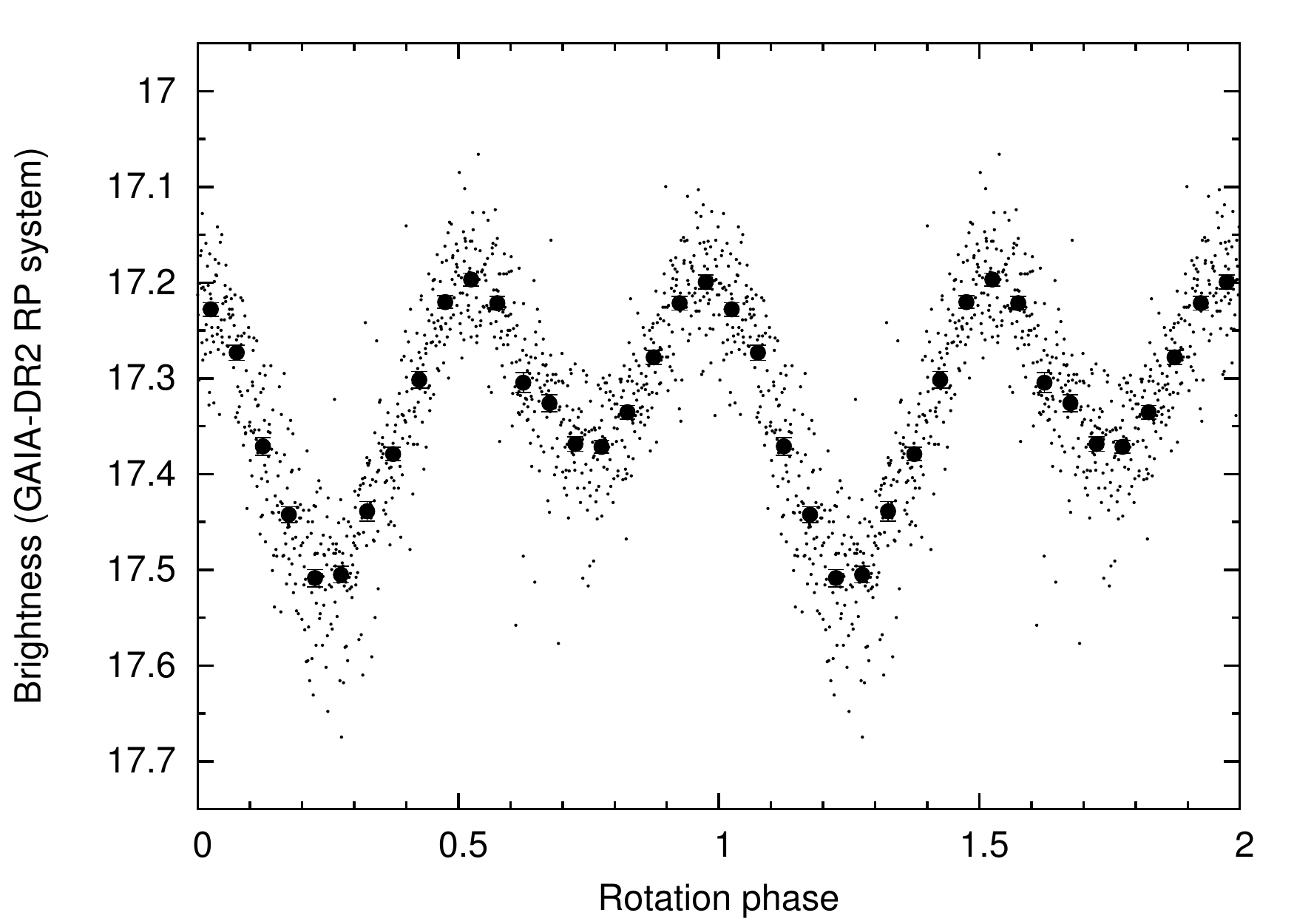}}%
\resizebox{57mm}{!}{\includegraphics{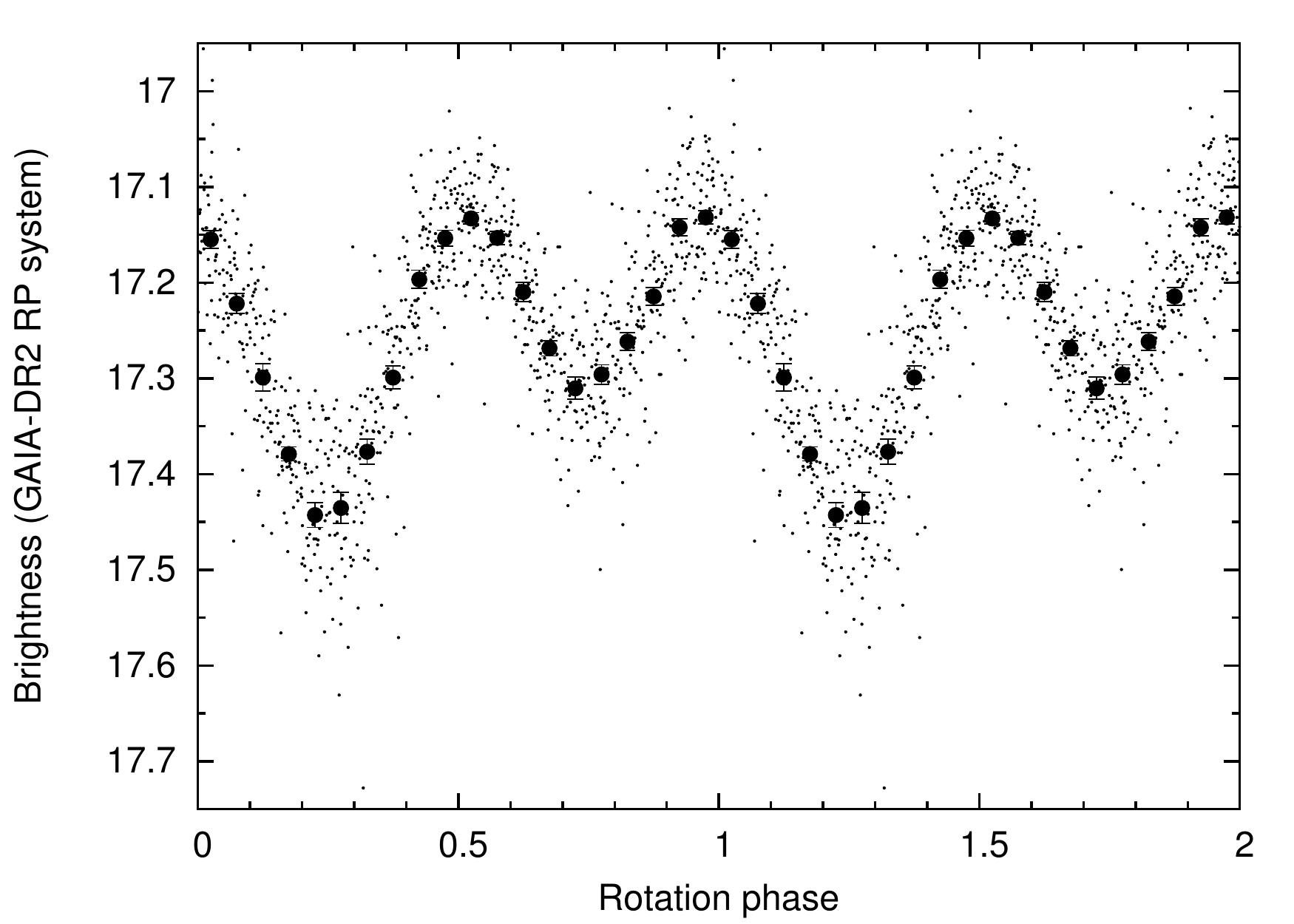}}%
\resizebox{57mm}{!}{\includegraphics{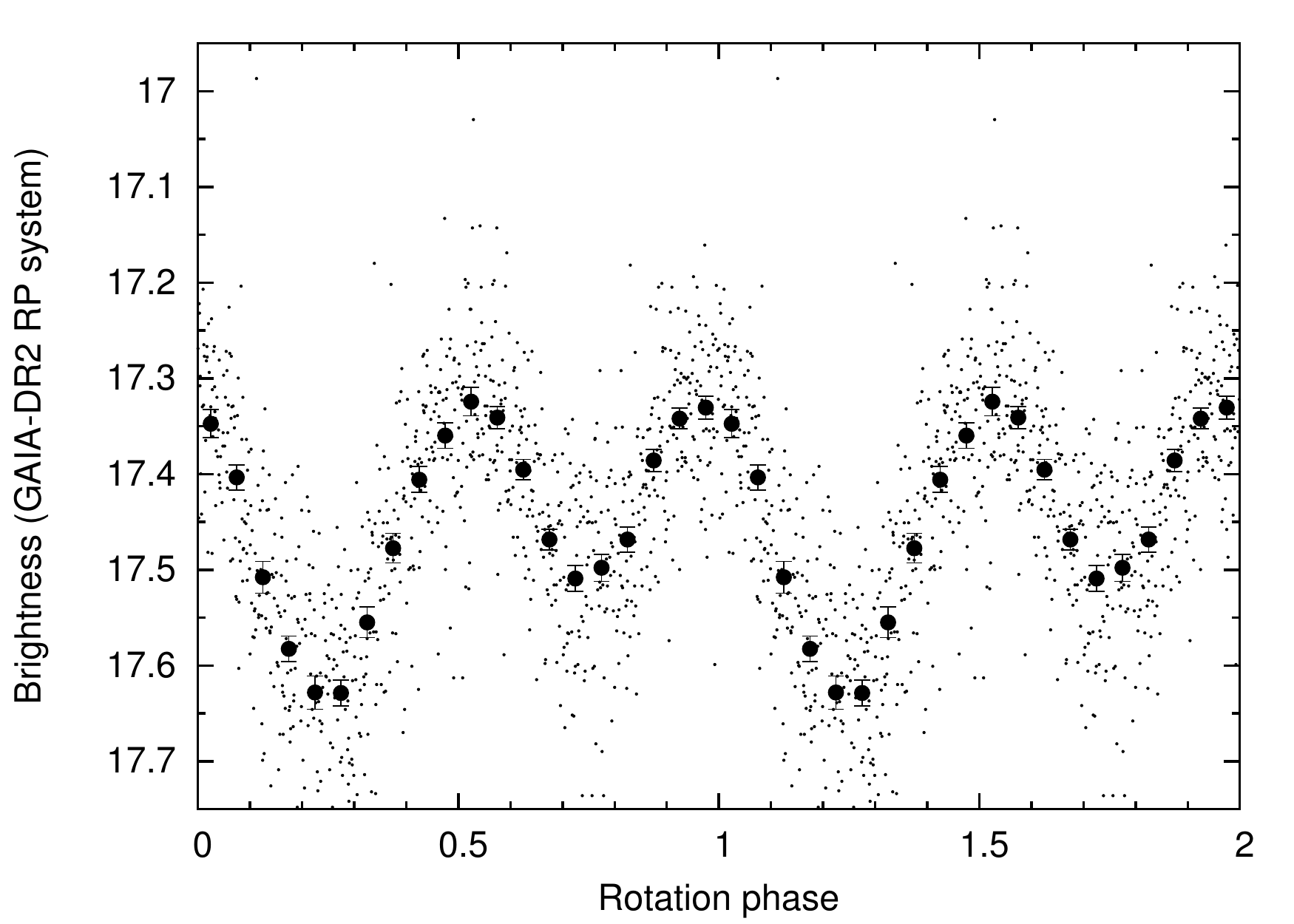}}%

\noindent
\resizebox{57mm}{!}{\includegraphics{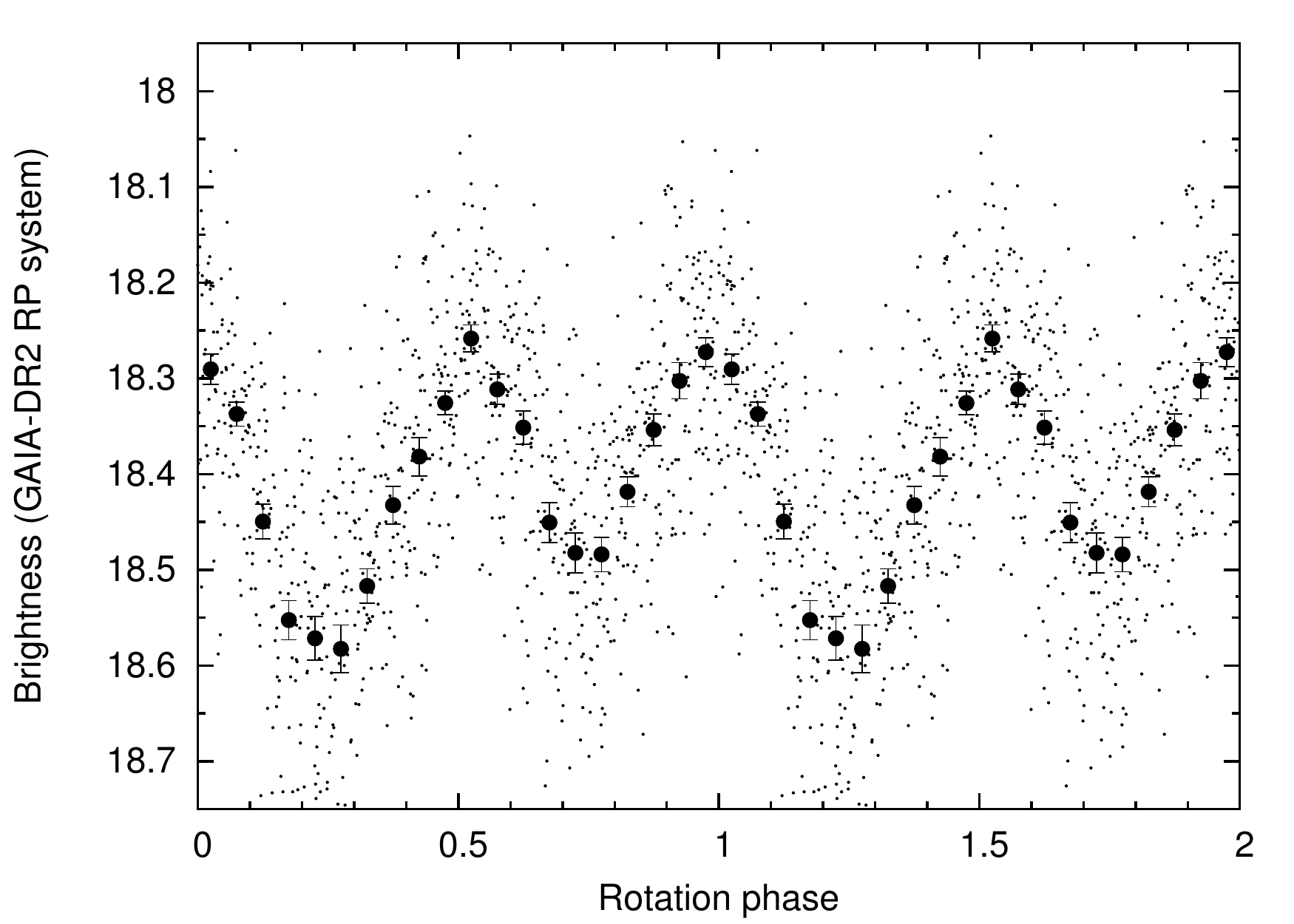}}%
\resizebox{57mm}{!}{\includegraphics{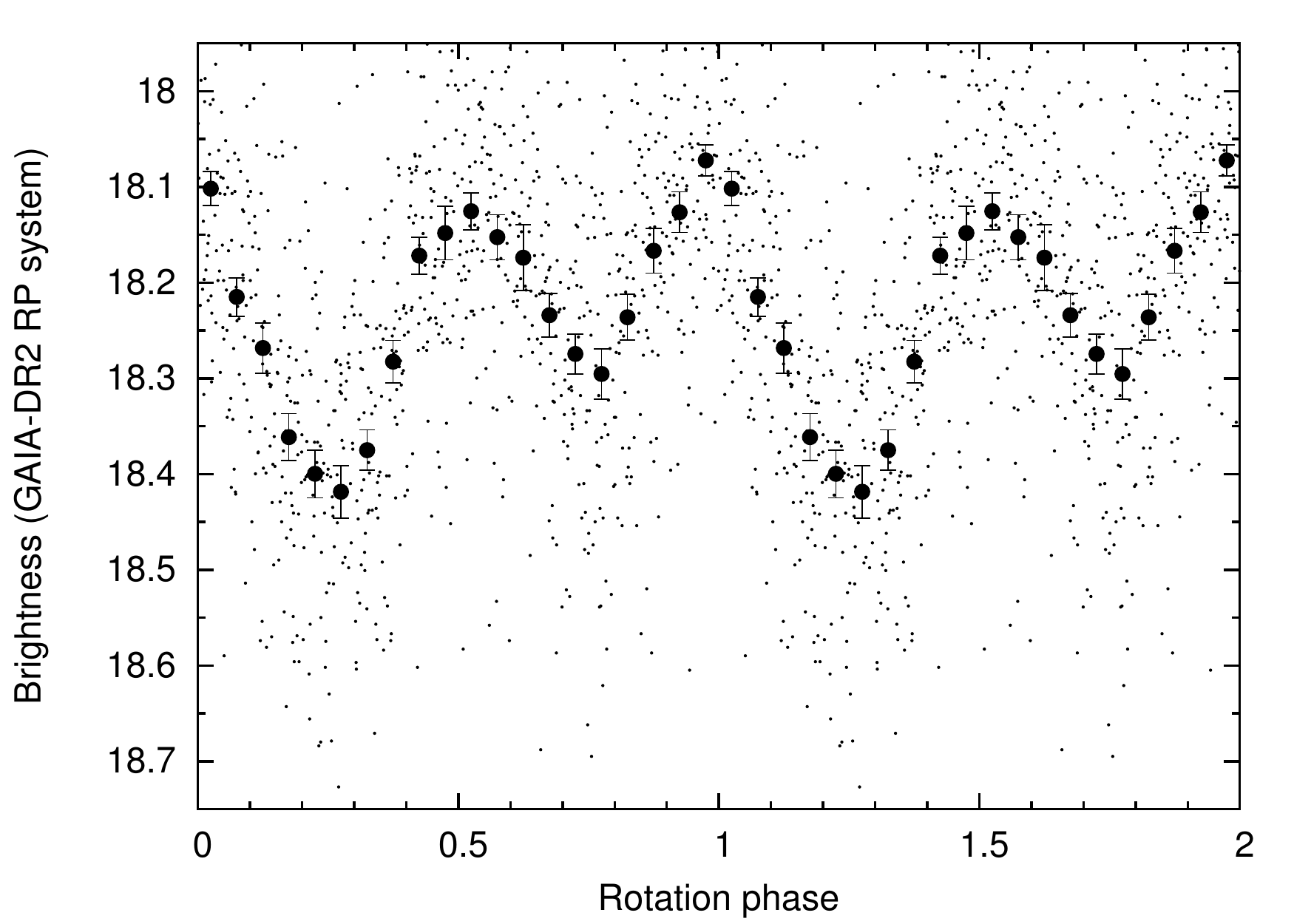}}%
\resizebox{57mm}{!}{\includegraphics{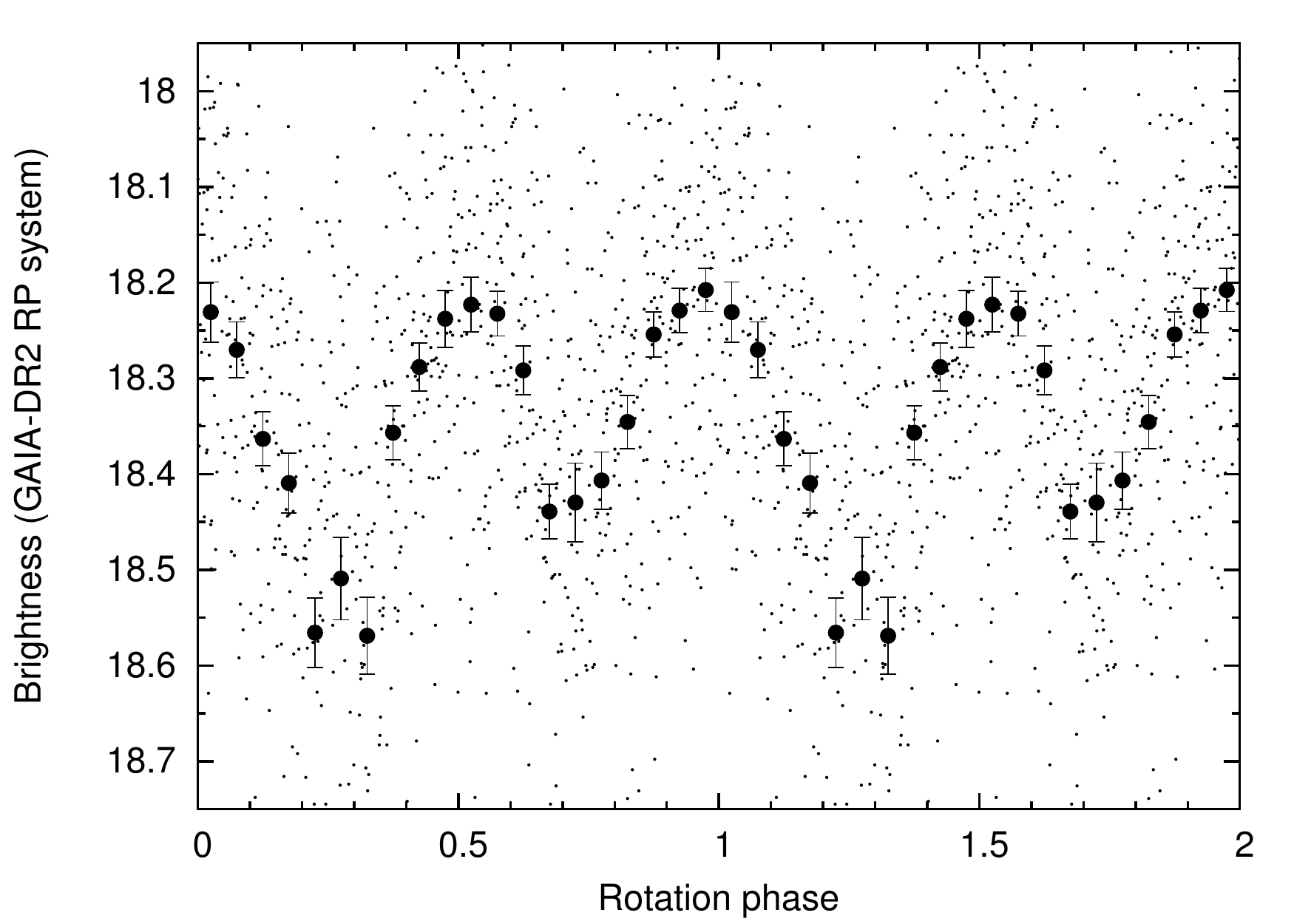}}%
\end{center}
\caption{Simulated light curves of real asteroids having fictitious 
rotational characteristics for various apparent brightness values and 
galactic latitudes. These $3\times4$ panels show light curves with 
double-peaked harmonic variations, having a period of $P=0.30792\,{\rm 
d}$ and a peak-to-peak amplitude of $0.3^{\rm m}$. The light curves are 
continuous and cover a time interval of $20^{\rm d}$, slightly shorter 
than an observing cycle of a TESS sector. The four rows correspond to 
mean ${\rm V}$ brightness of $16^{\rm m}$, $17^{\rm m}$, $18^{\rm m}$ 
and $19^{\rm m}$ while the three columns correspond to galactic 
latitudes of $|\beta|\approx 45^\circ$, $|\beta|\approx 20^\circ$ and 
$|\beta|\approx 10^\circ$, respectively. These simulations reveal the 
effect of the different stellar background densities. }
\label{fig:minors}
\end{figure*}


We should note here that due to the nature of this type of analysis, it 
is not essential to fully incorporate \emph{all} of the properties of 
the TESS system. Namely, here we ignored the spatial variations of the 
point-spread function \citep[which was taken into account in the 
reference simulations, see][]{sullivan2015} and used a simple Gaussian 
PSF model having a full width at half maximum (FWHM) corresponding to 
the respective ensquared energy \citep[see Table~1 in ][]{ricker2015}. 
In addition, the large-scale optical distortions were also neglected. 
This is mainly due to the fact that signal-to-noise ratios of (fainter) 
minor planets will predominantly be determined by the confusion of 
background stars -- and this is a much more local effect than the 
large-scale optical distortions of the TESS optics. The light curves are 
generated using differential photometry where subtracted images were 
derived using the corresponding FITSH tools (\texttt{fiarith}, 
\texttt{ficonv}). If the spacecraft pointing jitter is negligible, 
simple per-pixel arithmetics are sufficient while in the case of large 
jitter (even when it contains non-white noise components), image convolution 
is the most efficient way to retrieve the differential images and use 
the convolution kernels themselves to optimize the aperture \citep[see 
Eqs.~80-83 in][]{pal2009}. The reference image for differential 
photometry has been chosen as the median for the given time series.

In the next section, we use these tools to simulate both the light 
curves of minor planets from artificial 
TESS images as well as the effects of minor planet encounters on the 
photometry of target stars. Statistics and expectations about the 
occurrence rates of such minor planet encounters are presented in the 
following section.


\begin{figure*}
\begin{center}
\resizebox{160mm}{!}{\includegraphics{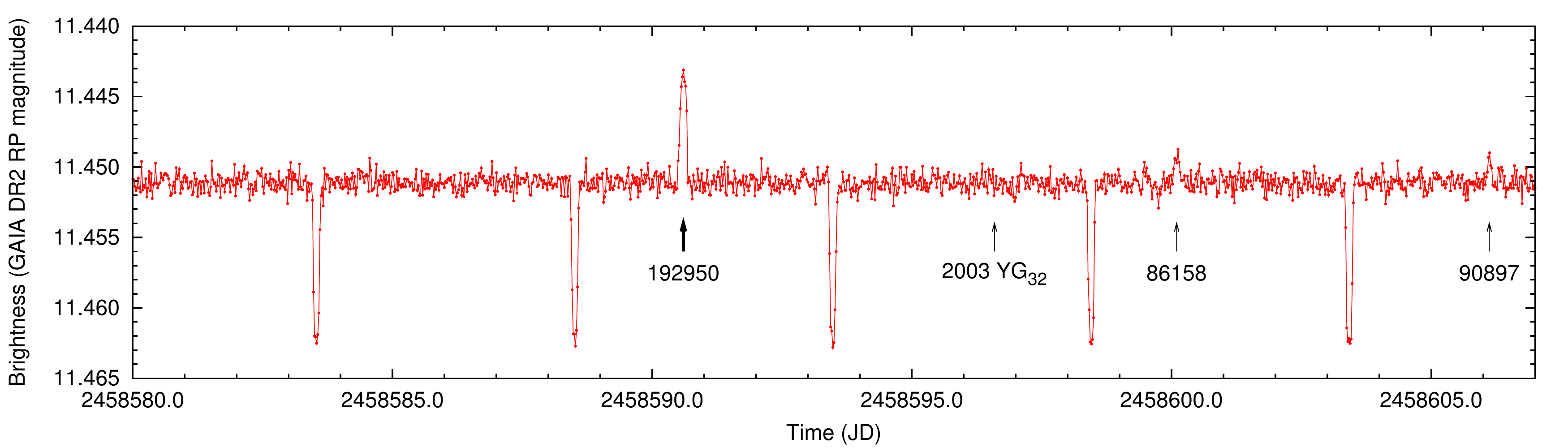}}
\end{center}
\caption{Simulation of the light curve of the transiting exoplanet
host star WASP-83 between JDs 2458580 and 2458607 (Apr 6.5, 2019 and 
May 3.5, 2019, respectively), in accordance with the expected scheduling
of the primary TESS mission. During this period, the target star is 
encountered by four minor planets where (192950) 2000\,BQ$_{2}$ is the one with
the highest expected brightness of $\sim 16.7^{\rm m}$ in the TESS passbands (assuming
a grey color). The encounter with (192950) 2000\,BQ$_{2}$ would yield an excess
in the total brightness around $\sim 8$ millimagnitudes, comparable
with the transit depth. In addition, the apparent encounters with
(86158) 1999\,RY$_{203}$ and (90897) 1997\,CF$_{6}$ will also yield excesses
around $\sim 2$\,mmag where the actual objects have an apparent magnitude
of $\sim 18.1^{\rm m}$  in the TESS passbands. The faintest one, 2003\,YG$_{32}$
with an apparent brightness of $\sim 21^{\rm m}$ yields no significant excess.}
\label{fig:wasp83lc}
\end{figure*}

\section{Light curve simulations}
\label{sec:lcsimulations}

\subsection{(136199) Eris as a target dwarf planet}
\label{sec:eris}

(136199) Eris is a bright dwarf planet, the most distant known natural
object in the Solar System. Recent studies \citep{sicardy2011,santossanz2012}
revealed in an unambiguous manner that its surface is extremely reflective
and up to now, no solid rotation period has been estimated for this planet.
Since Eris currently has an ecliptic latitude of 
$\beta\approx -12^\circ$, it cannot be observed in the K2 mission but it will be
an ideal target for TESS Camera \#1. The expected brightness of 
(136199) Eris is around $\sim18.1^{\rm m}$ in the TESS passbands. 
By performing the procedures presented in Sec.~\ref{sec:tools}, we generated
a 27-days long light curve of Eris, starting from JD\,2458411 and ending
at JD\,2458438. Like any of the small Solar System bodies, Eris has a 
retrograde motion during this period while the total span of the object
throughout such a TESS campaign of 27 days is approximately $15^\prime$.
Therefore, one can expect that Eris is a point-like source since this
aforementioned motion is equivalent to $\sim 0.03$ TESS pixels per 
long cadence.

In order to test the expectations of TESS photometric yield, we injected 
a sinusoidal single-peaked variation in the brightness of Eris, having a 
peak-to-peak amplitude of $0.05^{\rm m}$ and a period corresponding to 
the known orbital period of its moon, Dysnomia 
\citep[15.77\,d,][]{Brown+Schaller}. In Fig.~\ref{fig:eris} we 
demonstrate that the rotation period with this amplitude can safely be 
recovered in the frequency space, and the folded light curve is also 
prominent.

\subsection{Main-belt asteroids as target bodies}
\label{sec:minor}

In order to have an insight about the expected photometric quality of 
main-belt minor planets, we carried out a series of simulations by 
implanting variations in the light curves of known bodies. We selected 
three regions in the sky with higher, intermediate and low galactic 
latitudes of $|\beta|\approx 45^\circ$, $|\beta|\approx 20^\circ$ and 
$|\beta|\approx 10^\circ$ and queried for $4+4+4$ objects having a mean 
visual brightness of $\sim 16^{\rm m}$, $\sim 17^{\rm m}$, $\sim 18^{\rm 
m}$ and $\sim 19^{\rm m}$. The implanted variations were generated with 
a periodicity of $P=0.30792\,{\rm days}$ and with a peak-to-peak 
amplitude of $0.3^{\rm m}$. Here this period of $P=0.30792\,{\rm 
d}=7.39\,{\rm h}$ corresponds to the median rotation period of the 
main-belt asteroids as found in \cite{warner2009}. The signal itself was 
assumed to be a slightly asymmetric double-peaked harmonic function for 
all cases. The resulting $3\times4=12$ light curves are displayed on 
Fig.~\ref{fig:minors}. Note that the mean brightness in the TESS/Gaia 
($G_{\rm RP}$) passbands are higher by $\approx 0.7^{\rm m}$ due to the 
intrinsic color of the Sun. As it can be seen, using these $\sim 
20\,{\rm d}$ long signals, the light curve shape can be reconstructed 
for all of the cases and the effect of the increasing stellar background 
is also prominent as we go for lower and lower galactic latitudes.

\subsection{WASP-83 as a target star}
\label{sec:wasp83}

In this section, we present a simulated light curve of WASP-83, an 
$G_{\rm RP} = 11.45^{\rm m}$, late G-type star orbited by a $\sim 
1\,R_{\rm Jup}$ planet \citep{hellier2015}. According to the expected 
schedule of TESS, this planet host star will be observed around the 
second half of April 2019.

Due to the low ecliptic latitude, $\beta \approx -13.7^\circ$, this star 
has a quite high chance of nearby encounters by small, foreground Solar System 
bodies (see e.g. Fig.~\ref{fig:segmentdrawing}, the expected density of 
minor planets is around 40\% of the peak density). Throughout our 
simulations, we expected that the corresponding TESS sector is going to 
be observed between April 6 and May 3, 2019. By exploiting EPHEMD (see 
Sec.~\ref{sec:tools} above), we found that within this period, four 
minor planets would encounter within 0.02 degrees, i.e. within $\lesssim 
3$ TESS pixels where one of the objects has a brightness of $\sim$ 1\% 
of the target star. Indeed, the simulated light curve displayed in
Fig.~\ref{fig:wasp83lc} clearly features flux excesses during intervals 
of several hours caused by objects brighter than $19^{\rm m}$.


\section{Encounter statistics}
\label{sec:encounters}

In the previous section (Sec.~\ref{sec:wasp83}) we demonstrated a case 
where the light curve of a known target star harboring a transiting 
exoplanet is affected by some minor planet encounters. In the following 
we present brief statistics about the expected number for such 
encounters, depending on the ecliptic latitudes. In this experiment, we 
created $36\times 7$ sub-fields, each having a diameter of $1$ degree and 
distributed uniformly on a grid with a spacing of $10^\circ$ along the 
ecliptic between $\beta=-30^\circ$ and $\beta=+30^\circ$. For each 
field, we retrieved the \textit{Gaia} DR2 catalogue down to $17^{\rm m}$ 
in the $G_\mathrm{RP}$ magnitude and computed the properties of minor 
planet encounters (within 2 TESS pixels). These encounters were queried 
during a fictitious TESS sector observation run, lasting for $27$ days and 
centered at the corresponding anti-Sun longitude at the mid-time of the 
campaign.

We found that the average number of encounters per star do not depend 
significantly on the ecliptic latitude, namely for $\beta=\pm30^\circ$, 
the number of such events is $0.84\pm0.18$, for $\beta=\pm20^\circ$, it 
is $1.52\pm0.22$, for $\beta=\pm10^\circ$ it is $3.67\pm0.58$ and for 
the ecliptic itself it is $6.93\pm0.64$. Considering only the encounters 
which would yield at least 1\% flux increment (i.e., comparable to the 
transit depth caused by a hot Jupiter), the numbers are somewhat less, 
i.e. $0.70\pm0.16$, $1.31\pm0.17$, $3.07\pm0.43$ and $5.84\pm0.47$, for 
$\beta=\pm30^\circ$, $\pm20^\circ$, $\pm10^\circ$ and $0^\circ$, 
respectively. If we are interested in stars which are brighter than 
$13^{\rm m}$ in the TESS passbands, then the number of encounters 
yielding at least 1\% flux increment is around $\sim 0.10$, $\sim 0.20$, 
$\sim 0.39$ and $\sim 0.62$ for $\beta=\pm30^\circ$, $\pm20^\circ$, 
$\pm10^\circ$ and $0^\circ$, respectively. All in all, we conclude by 
comparing the aforementioned numbers with the the light curve of WASP-83 
(see Fig.~\ref{fig:wasp83lc}) and knowing that this star has an ecliptic 
latitude of $\beta \approx -13.7^\circ$, that this star can be 
considered as a rather ``typical case'' regarding the expected 
effects of minor planet encounters on the light curve.


\section{Summary}
\label{sec:summary}

Similar to the {\it Kepler}/K2 mission, Solar System objects can be 
considered either as targets of observations or noise and/or flux excess 
sources for stars. We found that in each campaign, rather good 
photometry can be achieved for some thousands of minor planets. Light 
curves are uninterrupted for several dozens of rotations and basic 
physical characteristics can be derived for objects with very long 
rotational periods as well -- without any ambiguity. While most of the 
trans-Neptunian objects are too faint for TESS, the brighter ones can 
also be targets of interest. Most notably, the dwarf planet Eris will 
also fall on silicon during the primary mission. Eris is bright enough 
for TESS photometry and moves on a field which is not significantly 
confused by stars. We also found that minor planets affect the 
photometric aperture for several hours due to the lower spatial 
resolution. Here we note that these tools of ours can also be applied to 
wide-field ground-based surveys, including full-sky monitoring such as 
in the KELT \citep{pepper2007}, Fly's Eye \citep{pal2013,pal2016b}, 
Evryscope \citep{law2015} or MASCARA \citep{snellen2012} 
initiatives.

We emphasize that extended mission setups of the TESS spacecraft can 
have versatile pointing configurations where the vicinity of the 
ecliptic can be covered more effectively -- i.e., filling the gap at 
$|\beta|\lesssim 6^\circ$, between the northern and southern primary 
survey areas and/or filling the gaps between the prime mission sectors 
at $6^\circ\lesssim |\beta|\lesssim 30^\circ$, see also 
\cite{bouma2017}. Such configurations will yield not only important 
follow-up observations for the K2 mission but also much more 
possibilities for Solar System studies. Other proposals, however, argue 
for increasing the time coverage and/or the size of the continuous 
viewing zone at the expense of the total spatial coverage \citep[see 
e.g. the ``Camera 3 on the ecliptic pole'', C3PO configuration of][]{huang2018}. 
These setups yield field-of-view only at 
higher ecliptic latitudes (e.g. $30^\circ\lesssim|\beta|$ for C3PO or 
$42^\circ\lesssim|\beta|$ when the spacecraft reference pointing is set 
to the ecliptic poles), resulting in significantly lesser number of 
observations for Solar System objects.

\acknowledgments 
This work has been supported by the ``Lend\"ulet'' 
grants LP2012-31 and LP2018-07 of the Hungarian Academy of Sciences, and 
by the Hungarian National Research, Development and Innovation Office 
(NKFIH) grants PD-116175 and K-125015. A.P. acknowledges the MIT Kavli 
Center and the Kavli Foundation for their hospitality during the stays 
at MIT and the NASA contract number NNG14FC03C. L.M. was supported by 
the J\'anos Bolyai Research Scholarship of the Hungarian Academy of 
Sciences. The research leading to these results has also received 
funding from the European Unions Horizon 2020 Research and Innovation 
Programme, under Grant Agreement no 687378 "Small Bodies: Near and Far". 
This work has made use of data from the European Space Agency (ESA) 
mission {\it Gaia} (\url{https://www.cosmos.esa.int/gaia}), processed by 
the {\it Gaia} Data Processing and Analysis Consortium (DPAC, 
\url{https://www.cosmos.esa.int/web/gaia/dpac/consortium}).  Funding for 
the DPAC has been provided by national institutions, in particular the 
institutions participating in the {\it Gaia} Multilateral Agreement.

{}


\end{document}